\documentclass[fleqn,usenatbib]{mnras}

\usepackage{newtxtext,newtxmath}
\usepackage{booktabs}

\usepackage[T1]{fontenc}

\DeclareRobustCommand{\VAN}[3]{#2}
\let\VANthebibliography\thebibliography
\def\thebibliography{\DeclareRobustCommand{\VAN}[3]{##3}\VANthebibliography}

\usepackage{graphicx}	
\usepackage{amsmath}	

\title[Hubble tension problem in the HDE model]{Revisiting the Hubble tension problem in the framework of holographic dark energy}

\author[J.-X. Li and S. Wang]{
Jun-Xian Li,$^{1}$\thanks{E-mail: lijx389@mail2.sysu.edu.cn} Shuang Wang,$^{1}$\thanks{E-mail: wangshuang@mail.sysu.edu.cn (corresponding author)}
\\
$^{1}$School of Physics and Astronomy, Sun Yat-sen University, No.2 Daxue Road, Tangjiawan, Zhuhai, 519082, P.R.China\\
}

\date{Accepted XXX. Received YYY; in original form ZZZ}

\pubyear{\the\year{}}

\begin{document}
\label{firstpage}
\pagerange{\pageref{firstpage}--\pageref{lastpage}}
\maketitle

\begin{abstract}

The Hubble tension problem is one of the most significant challenges in modern cosmology.
In this paper, we study the Hubble tension problem in the framework of holographic dark energy (HDE). 
To perform a systematic and comprehensive analysis, we select six representative theoretical models from all four categories of HDE.
For the observational data, we adopt the Baryon Acoustic Oscillation (BAO) data from the Dark Energy Spectroscopic Instrument (DESI) Data Release 2 (DR2) along with a collection of alternative BAO measurements, Cosmic Microwave Background (CMB) distance priors from $Planck$ 2018, and type Ia supernovae (SN) data from the PantheonPlus, Union3, and DESY5 compilations. 
We find that HDE models that employ the Hubble scale or its combinations as the infrared (IR) cutoff cannot alleviate the Hubble tension problem. 
In contrast, HDE models that employ the future event horizon as the IR cutoff can partially mitigate the Hubble tension problem.
It must be stressed that these two key conclusions hold true for cases of adopting different theoretical HDE models and different observational data. 
Our findings advocate for further exploration of HDE models using other types of cosmological observations.

\end{abstract}





\section{Introduction}


As the standard model of cosmology, the $\Lambda$ Cold Dark Matter ($\Lambda$CDM) model has provided a good description of a range of observational data.
However, in recent years, this standard model has faced increasing observational challenges.
The most pressing issue is the Hubble tension problem, a discrepancy exceeding 5$\sigma$ confidence level (CL) between the directly measured current cosmic expansion rate and its inferred value from early-universe observations (see \cite{Poulin:2023lkg, Rong-Gen:2023dcz, Vagnozzi:2023nrq, Efstathiou:2024dvn, CosmoVerseNetwork:2025alb} for reviews).
Specifically, a recent direct measurement from the SH0ES Collaboration gives $H_0^{SH0ES} = 73.17 \pm 0.86$ km s$^{-1}$ Mpc$^{-1}$ \citep{Riess:2021jrx, Breuval:2024lsv}; this value differs significantly from the indirect inference value, obtained from $Planck$ Cosmic Microwave Background (CMB) measurements, which gives $H_0^{Planck} = 67.4 \pm 0.5$ km s$^{-1}$ Mpc$^{-1}$ \citep{Planck:2018vyg}.
In fact, the essence of the Hubble tension can be viewed as a cosmological trilemma: the standard $\Lambda$CDM model, early-time measurements (e.g. the CMB), and late-time measurements (e.g. the local distance ladder) cannot be simultaneously correct \textemdash{} implying that at least one element of this triad is incorrect or incomplete. 
If one assumes the reliability of current observational data, it is necessary to investigate other cosmological models beyond the $\Lambda$CDM model.

In addition to the Hubble tension problem, recent results from the Dark Energy Spectroscopic Instrument (DESI) collaboration \citep{DESI:2024mwx, DESI:2025zgx} provided further evidence for deviation from $\Lambda$CDM. 
Using the combination of DESI Data Release 2 (DR2) and CMB data, the DESI found a 3.1$\sigma$ CL preference for the Chevallier-Polarski-Linder (CPL) model \citep{Chevallier:2000qy, Linder:2002et} over $\Lambda$CDM.
Moreover, after adding the PantheonPlus \citep{Brout:2022vxf}, Union3 \citep{Rubin:2023jdq}, or DESY5 \citep{DES:2024jxu} type Ia supernovae (SN) dataset into the combination, the preference is 2.8$\sigma$, 3.8$\sigma$ or 4.2$\sigma$, respectively.
These deviations from the cosmological constant have sparked extensive debates on dynamical dark energy (DE), including
scalar field DE models \citep{Yin:2024hba, Shlivko:2024llw, DESI:2024kob, Notari:2024rti, Wolf:2024eph, Anchordoqui:2025fgz, Wang:2025znm, Wolf:2025acj}, interacting DE models \citep{Giare:2024smz, Montani:2024pou, Li:2024qso, Li:2025owk, Li:2025ula, Yang:2025vnm, Silva:2025hxw, Zhai:2025hfi, Feng:2025mlo, Li:2025dwz, Li:2025muv, Zhang:2025dwu, Pan:2025qwy, Yang:2025boq}, early DE models \citep{ Qu:2024lpx, Seto:2024cgo, Chaussidon:2025npr}, phenomenological DE models \citep{Kessler:2025kju, Specogna:2025guo, Du:2025xes, Yang:2025oax, Cheng:2025lod}, and some others \citep{Wang:2024rus, RoyChoudhury:2024wri, RoyChoudhury:2025dhe, Pang:2024wul, Huang:2025som, RoyChoudhury:2025iis, Jiang:2024viw, Pang:2025lvh, Li:2025ops} \footnote{For recent discussions on modified gravity model, we refer the reader to Refs. \citep{Yang:2024kdo, Yang:2025mws, Li:2025cxn, Pan:2025psn, Feng:2025cwi}.}. For related reviews, we refer the reader to Refs. \citep{Yoo:2012ug,  Arun:2017uaw, Bahamonde:2017ize, Cai:2025mas}.

As a prominent class of dynamical DE models, holographic dark energy (HDE) originates from a theoretical attempt of applying the holographic principle (HP) \citep{tHooft:1993dmi, Susskind:1994vu} to the DE problem. The HP states that all information contained within a volume of space can be encoded on its boundary, much like a hologram. This implies that the energy density of DE $\rho_{de}$ can be described by quantities on the boundary of the universe, including the characteristic length scale $L$ (i.e. infrared (IR) cutoff) and the reduced Planck mass $M_P^2 \equiv 1/(8\pi G)$. 
Based on the dimensional analysis, we have \citep{Wang:2016och}:
\begin{equation}
\label{rhode-exp}
    \rho_{de} = C_1 M_P^4 + C_2 M_P^2 L^{-2} + C_3 L^{-4} + \cdots,
\end{equation}
where $C_1$, $C_2$, $C_3$ are constant parameters. However, the leading term is $10^{120}$ times larger than the cosmological observations \citep{Peebles:2002gy}, so this term should be excluded. Moreover, the third term and the other terms are negligible compared to the second term. Therefore, the expression of $\rho_{de}$ can be written as \citep{Wang:2016och}:
\begin{equation}
\label{rhode}
    \rho_{de} = 3C^2 M_P^2 L^{-2},
\end{equation}
where $C$ is a dimensionless constant parameter. It is important to emphasize that Eq. (\ref{rhode}) serves as the foundational expression for the HDE energy density.
Over the past two decades, extensive studies on HDE have led to various theoretical models \footnote{For related approaches to HDE generalization, we refer the reader to \cite{Nojiri:2017opc, Nojiri:2019kkp, Nojiri:2020wmh, Nojiri:2021iko, Nojiri:2021jxf}.}. 
As pointed out by \cite{Wang:2023gov}, HDE models can be divided into four main categories:
(1) HDE models with other characteristic length scale \citep{Li:2004rb, Cai:2007us, Wei:2007ty, Huang:2012xma}; 
(2) HDE models with extended Hubble scale \citep{Nojiri:2005pu, Gao:2007ep, Granda:2008dk, Gong:2009dc}; 
(3) HDE models with dark sector interaction \citep{Zimdahl:2007zz, Wang:2005jx, Setare:2006wh, Wei:2007ut, Xu:2009ys}; 
(4) HDE models with modified black hole entropy \citep{Tavayef:2018xwx, Saridakis:2018unr, Saridakis:2020zol, Srivastava:2020cyk, Adhikary:2021xym, Drepanou:2021jiv}.
In addition to the proposal of theoretical models, numerical studies on HDE models have also attracted considerable attention \citep{Li:2009bn, Li:2009zs, Zhang:2012qra, Li:2012spm, Wang:2013zca, Wen:2017aaa}. 
For recent progress, we refer the reader to Refs. \citep{Tyagi:2024cqp, Astashenok:2024jje, Brevik:2024ozg, Li:2024qus, Han:2024sxm, Feng:2025wbz, Cimdiker:2025vfn, Luciano:2025elo, Zhang:2025oki, Ren:2025jhe, Guin:2025xki, Wu:2025vfs, Shen:2025cjm}.

In the previous literature, many attempts have been made to use HDE models to discuss the Hubble tension problem.
For example, using $Planck$ 2015 data and baryon acoustic oscillation (BAO) data, \cite{Zhao:2017urm} incorporated the effective mass of the sterile neutrino ($m^{\text{eff}}_{\nu,\text{sterile}}$) and the effective number of relativistic species ($N_{\text{eff}}$) into the original HDE model, obtaining $H_0=72.4^{+1.9}_{-2.2}$ km s$^{-1}$ Mpc$^{-1}$.  \cite{Guo:2018ans} investigated the original HDE model with $m^{\text{eff}}_{\nu,\text{sterile}}$ using $Planck$ 2015 data, BAO data, JLA SN data, and local measurement of $H_0$ data, finding that the Hubble tension could be reduced to 1.11$\sigma$. 
Using $Planck$ 2018 CMB data and BAO data, \cite{Dai:2020rfo} showed that the original HDE model reconciles the Hubble tension to 1.4$\sigma$.
More recently, \cite{Tang:2024gtq} examined the original HDE model with $N_{\text{eff}}$ and found that the model brings down the Hubble tension to 0.85$\sigma$, using $Planck$ and Atacama Cosmology Telescope (ACT) CMB data, BAO and redshift-space distortion (RSD) data, and the measurement from the SH0ES team.

It should be stressed that previous studies only focused on some individual HDE models.
In order to obtain a comprehensive understanding of HDE, it is necessary to perform a systematic study for all four categories of HDE models.
For example, in our recent paper \citep{Li:2024bwr}, we performed a comprehensive numerical analysis of all four categories of HDE models using DESI DR1 data.
By comparing different HDE models, we found that the choice of the IR cutoff plays an important role in confronting cosmological data.
Through systematic analysis, one can identify the general features of HDE models that provide better fits to the data.
Therefore, in this work,  we systematically investigate the Hubble tension problem based on all four categories of HDE models.
Specifically, we select one or two representative models from each category, as summarized in Table \ref{tab:hde_models}. 
For the first category, we consider the original HDE (OHDE) model \citep{Li:2004rb}. 
For the second category, we analyze the generalized Ricci HDE (GRDE) model \citep{Granda:2008dk}. 
For the third category, we examine two interacting HDE models: IHDE1 \citep{Zimdahl:2007zz}, which uses the Hubble scale as the IR cutoff, and IHDE2 \citep{Wang:2005jx}, which uses the future event horizon as the IR cutoff. 
For the fourth category, we study the Tsallis HDE (THDE) model \citep{Tavayef:2018xwx} and Barrow HDE (BHDE) model \citep{Saridakis:2020zol}, where THDE employs the Hubble scale as the cutoff and BHDE employs the future event horizon. 
Therefore, in this paper, we use six representative HDE models to investigate the Hubble tension problem.
We also include the $\Lambda$CDM model as a fiducial benchmark for comparison.
For observational data, we adopt the latest BAO data from DESI DR2, CMB distance priors from $Planck$ 2018, and SN data from DESY5 compilation.
To assess the robustness of our results, we also compare with alternative datasets including a collection of previous BAO measurements, as well as the PantheonPlus and Union3 SN compilations.

\begin{table*}
\begin{center}
\begin{tabular}{llll}
\hline \hline
$\textbf{Model Category}$ & $\textbf{Model}$ & $\textbf{Length scale $L$}$ & $\textbf{Reference}$ \\
\hline \hline
{HDE models with other characteristic length scale}
  &  OHDE   & future event horizon   &  \cite{Li:2004rb}  \\
\hline
{HDE models with extended Hubble scale}
  &  GRDE   & extended Hubble scale  &  \cite{Granda:2008dk}\\
\hline
{HDE models with dark sector interaction}
  &  IHDE1  & Hubble scale &  \cite{Zimdahl:2007zz} \\
  &  IHDE2  & future event horizon &  \cite{Wang:2005jx} \\
\hline
{HDE models with modified black hole entropy}
  &  THDE  & Hubble scale  &  \cite{Tavayef:2018xwx} \\
  &  BHDE  & future event horizon &  \cite{Saridakis:2020zol} \\
\hline
\end{tabular}
\end{center}
\caption{Four categories of HDE models. We list the models considered in this paper along with their characteristic lengths and references.}
\label{tab:hde_models}
\end{table*}

The structure of this paper is as follows: Section \ref{sec:data_analysis} describes the observational datasets and the analysis framework. 
Section \ref{sec:HDE} introduces the representative HDE models and presents the corresponding cosmology-fits. 
Section \ref{sec:results} discusses the Hubble tension problem in the framework of HDE. 
Finally, Section \ref{sec:conclusion} gives a summary. 
Throughout this paper, the spatially flat FLRW universe is adopted and the subscript '0', denoted as present day, is dropped on the density parameters.

\section{Datasets and methodology}
\label{sec:data_analysis}
In this section, we first introduce each observational data from subsection \ref{subsec:bao} to subsection \ref{subsec:sn}. And then we describe the analysis methodology in subsection \ref{subsec:chisq}.

\subsection{Baryon acoustic oscillation}
\label{subsec:bao}
For the BAO data, we use two datasets.
The first is the latest DESI DR2 BAO data; the second is a collection of previous BAO measurements (hereafter non-DESI). In the following, we briefly introduce the composition of these two datasets.

The DESI DR2 BAO dataset \citep{DESI:2025zgx} comprise measurements from different classes of extragalactic targets: the bright galaxy sample (BGS) \citep{Hahn:2022dnf}, luminous red galaxies (LRG) \citep{DESI:2022gle}, emission line galaxies (ELG) \citep{Raichoor:2022jab}, and quasars (QSO)  \citep{Chaussidon:2022pqg}. 
Table \ref{tab:desi_BAO} lists the tracers, effective redshifts, observables, and measurement values for seven BAO data points. 
The covariance matrices are calculated using the coefficient $r$, and the detailed procedure can be found in \cite{Li:2024bwr}.
In the following text and figures, we refer to DESI DR2 simply as DESI.

\begin{table*}
\begin{center}
\renewcommand{\arraystretch}{1.2}
    \begin{tabular}{c|c|c|c|c}
    \hline \hline
    \textbf{DESI Tracer} & \textbf{$z_{\text{eff}}$} & \textbf{$D_M / r_d$} & \textbf{$D_H / r_d$} & \textbf{$r$ or $D_V / r_d$} \\
    \hline \hline
    BGS & 0.295 & --- & --- & 7.944 $\pm$ 0.075 \\
    LRG1 & 0.510 & 13.587 $\pm$ 0.169 & 21.863 $\pm$ 0.427 & $-0.475$ \\
    LRG2 & 0.706 & 17.347 $\pm$ 0.180 & 19.458 $\pm$ 0.332 & $-0.423$ \\
    LRG3+ELG1 & 0.934 & 21.574 $\pm$ 0.153 & 17.641 $\pm$ 0.193 & $-0.425$ \\
    ELG2 & 1.321 & 27.605 $\pm$ 0.320 & 14.178 $\pm$ 0.217 & $-0.444$ \\
    QSO & 1.484 & 30.519 $\pm$ 0.758 & 12.816 $\pm$ 0.513 & $-0.489$ \\
    Lya & 2.330 & 38.988 $\pm$ 0.531 & 8.632 $\pm$ 0.101 & $-0.431$ \\
    \hline
    \end{tabular}
\end{center}
\caption{Distance measurements from the DESI DR2 BAO data. We simply refer this dataset as $\textbf{DESI}$. Note that for each sample DESI measures either both $D_M / r_d$ and $D_H / r_d$, which are correlated with a coefficient $r$, or only $D_V / r_d$.}
\label{tab:desi_BAO}
\end{table*}

The non-DESI BAO dataset comprises measurements from 6-degree Field Galaxy Survey and Sloan Digital Sky Survey Main Galaxy Sample (6dFGS+SDSS MGS) \citep{Carter:2018vce}, Baryon Oscillation Spectroscopic Survey (BOSS) Galaxy \citep{eBOSS:2020hur}, extended BOSS (eBOSS) LRG \citep{eBOSS:2020hur, eBOSS:2020lta}, Dark Energy Survey Year 3 (DESY3) \citep{DES:2021wwk}, eBOSS Quasar \citep{eBOSS:2020gbb, eBOSS:2020uxp}, and eBOSS Ly$\alpha$-forest \citep{eBOSS:2020tmo}. 
Table \ref{tab:nondesi_BAO} lists the measurements, effective redshifts, observables, and measurement values for these seven BAO data points. 
The covariance matrices are provided in their respective references.
In the following text and figures, we refer to this BAO dataset simply as non-DESI.

\begin{table*}
\begin{center}
    \renewcommand{\arraystretch}{1.2}
    \begin{tabular}{c|c|c|c|c}
    \hline \hline
    \textbf{Measurements} & \textbf{$z_{\text{eff}}$} & \textbf{$D_M / r_d$} & \textbf{$D_H / r_d$} &  $D_V(r_{d,\text{fid}}/r_d)$ \\
    \hline \hline
    6dFGS+SDSS MGS & 0.122 & --- & --- & 539 $\pm$ 17 [Mpc]\\
    BOSS Galaxy & 0.38 & 10.274 $\pm$ 0.151 & 24.888 $\pm$ 0.582 & --- \\
    BOSS Galaxy & 0.51 & 13.381 $\pm$ 0.179 & 22.429 $\pm$ 0.482 & --- \\
    eBOSS LRG & 0.698 & 17.646 $\pm$ 0.302 & 19.770 $\pm$ 0.469 & --- \\
    DES Y3 & 0.835 & 18.92 $\pm$ 0.51 & --- & --- \\
    eBOSS Quasar & 1.48 & 30.21 $\pm$ 0.79 & 13.23 $\pm$ 0.47 & --- \\
    eBOSS Ly$\alpha$-forest & 2.334 & 37.5$^{+1.2}_{-1.1}$ & 8.99$^{+0.2}_{-0.19}$ & --- \\
    \hline
    \end{tabular}
\end{center}
    \caption{Distance measurements from different BAO measurements. We simply refer this dataset as $\textbf{non-DESI}$. Note that the sound horizon size of the fiducial model is $r_{d,\text{fid}}$ =147.5 Mpc.}
    \label{tab:nondesi_BAO}
\end{table*}

The quantities of BAO correspond to several key distances: $D_M$, $D_H$, and $D_V$. In a spatially flat FLRW universe, the transverse comoving distance $D_M$ at redshift $z$ is defined as
\begin{equation}
\label{eq.codist}
    D_M(z) = \frac{c}{H_0} \int_0^z \frac{d z'}{H(z') /H_0},
\end{equation}
where $c$ is the speed of light, $H_0=100h$ km $\text{s}^{-1} \text{Mpc}^{-1}$ is the present-day Hubble constant with $h$ being its dimensionless form. The distance variable $D_H$ is related to the Hubble parameter $H(z)$ as $D_H(z) = c/H(z)$. The angle-averaged distance $D_V$ is given by $D_V(z) = [zD_M(z)^2 D_H(z) ]^{1/3}$.

BAO measurements depend on the radius of the sound horizon at the drag epoch $r_d$. This represents the distance that acoustic waves can travel between the Big Bang and the drag epoch, which marks the time when baryons decoupled.
The sound horizon can be expressed as
\begin{equation}
\label{rs}
    r_s(z) = \int_{z}^\infty \frac{c_s(z')}{H(z')}dz',
\end{equation}
where $c_s(z)$ is the speed of sound. 
As noted in \cite{Brieden:2022heh, Schoneberg:2022ggi}, assuming standard pre-recombination physics, the drag epoch sound horizon can be computed by
\begin{equation}
    r_d \simeq \frac{147.05}{\mathrm{Mpc}} \left(\frac{\omega_\mathrm{m}}{0.1432}\right)^{-0.23} \left(\frac{\omega_b}{0.02236}\right)^{-0.13} \left(\frac{N_{\mathrm{eff}}}{3.04}\right)^{-0.1},
\end{equation}
where $\omega_m \equiv \Omega_m h^2$, $\omega_b \equiv \Omega_b h^2$, $\Omega_m$ and $\Omega_b$ denote the fractional energy densities of matter and baryons, and $N_{\mathrm{eff}}$ is the effective number of extra relativistic degrees of freedom. We assume standard neutrino content $N_{\mathrm{eff}} = 3.044$ in our analysis.

\subsection{Cosmic microwave background}
\label{subsec:cmb}
For the CMB data, we use the distance priors of $Planck$ 2018 from \cite{Zhai:2018vmm}. The method of distance priors \citep{Efstathiou:1998xx, Wang:2006ts, Wang:2007mza} compresses the full CMB data into the background quantities. This approach allows us to substitute the full CMB power spectrum with a more compact representation while retaining key cosmological information.

The distance priors contain two primary quantities: the shift parameter $R$ and the acoustic scale $l_a$. The shift parameter $R$ affects the peak heights in the CMB temperature power spectrum along the line of sight, while the acoustic scale $l_a$ influences the spacing of the peaks in the transverse direction. These parameters are defined as
\begin{align}
    R &\equiv \frac{ D_M(z_*)\sqrt{\Omega_m H_0^2}}{c},\\
    l_a &\equiv \frac{\pi D_M(z_*)}{r_s(z_*)},
\end{align}
where $z_*$ is the redshift at the photon decoupling epoch, which can be calculated by an approximate formula \citep{Hu:1995en}:
\begin{equation}
    z_* = 1048[1+0.00124(\Omega_b h^2)^{-0.738}][1+g_1(\Omega_m h^2)^{g_2}],
\end{equation}
where
\begin{align}
    g_1 &= \frac{0.0783(\Omega_b h^2)^{-0.238}}{1+39.5(\Omega_b h^2)^{0.763}},\\
    g_2 &= \frac{0.560}{1+21.1(\Omega_b h^2)^{1.81}}.
\end{align}
As pointed out by \cite{Wang:2007mza}, the baryon density $\Omega_b h^2$ should be included as an estimated parameter in the data analysis.
This is because its value is essential for computing the redshift $z_*$ at the photon decoupling epoch. 
Therefore, the CMB shift parameters $R$ and $l_a$ inherit an explicit dependence on $\Omega_b h^2$ through redshift $z_*$.

Since $\Omega_b h^2$ is correlated with $R$ and $l_a$, we should take into account their covariance. The data vector and its covariance matrix are the following respectively \citep{Zhai:2018vmm}:
\begin{equation}
\label{vetorV}
    V^{data} \equiv \begin{pmatrix}
    R \\
    l_a \\
    \Omega_b h^2 \\
    \end{pmatrix}
    =
    \begin{pmatrix}
    1.74963 \\
    301.80845 \\
    0.02237 \\
    \end{pmatrix},
\end{equation}

\begin{equation}
\begin{split}
    \text{Cov}_{\text{CMB}} = 10^{-8} \times 
    \begin{bmatrix}
    1598.9554 & 17112.007 & -36.311179 \\
    17112.007 &  811208.45 &  -494.79813 \\
    -36.311179 &  -494.79813 &  2.1242182
    \end{bmatrix}.
\end{split}
\end{equation}

It should be noted that in this work we use the CMB distance priors rather than the full CMB likelihood.
Compared with the distance priors approach, the use of full likelihood can provide tighter parameter constraints.
In addition, for models that modify late-time dynamics but may indirectly affect early-time quantities through parameter degeneracies, the full likelihood may help to break the degeneracies.
However, based on previous studies \citep{Zhai:2019nad, Boiza:2025xpn, Li:2025ops}, the difference between the distance priors and the full likelihood is small in cosmological parameter estimation.
Therefore, for simplicity and computational efficiency, we use CMB distance priors instead.

\subsection{Type Ia supernovae}
\label{subsec:sn}
For the SN data, we use three different compilations: PantheonPlus, Union3, and DESY5. 
The PantheonPlus (PPlus) compilation comprises 1550 supernovae, spanning a redshift range $0.01 \leq z \leq 2.26$ \citep{Brout:2022vxf}, which does not include the calibration sample at lower redshifts.
The Union3 compilation contains 2087 SN Ia over the redshift range $0.01<z<2.26$, which is processed through the Unity 1.5 pipeline based on Bayesian Hierarchical Modeling \citep{Rubin:2023jdq}.
The DESY5 compilation consists of 194 low-redshift SN Ia ($0.025<z<0.1$) and 1635 photometrically classified SN Ia covering the range $0.1<z<1.3$ \citep{DES:2024jxu}. 
All three likelihoods are implemented in the $\mathbf{Cobaya}$ sampling code and the underlying data are publicly available.\footnote{https://github.com/CobayaSampler/sn\_data.git} The covariance matrix, $\text{Cov}_{\text{SN}}$, which includes both statistical and systematic errors, is also publicly available.

SN data are usually specified in terms of the distance modulus $\mu$. The theoretical distance modulus $\mu_{th}$ in a flat universe is given by
\begin{equation}
    \mu_{th} = 5 \log_{10} \left[ \frac{d_L(z_{hel}, z_{cmb})}{\text{Mpc}} \right] +25,
\end{equation}
where $z_{hel}$ and $z_{cmb}$ are the heliocentric and CMB rest-frame redshifts of SN. The luminosity distance $d_L$ is
\begin{equation}
    d_L(z_{hel}, z_{cmb})= (1+z_{hel}) r(z_{cmb}),
\end{equation}
where $r(z)$ is the comoving distance given by Eq. (\ref{eq.codist}).

\subsection{Chi-square statistic}
\label{subsec:chisq}
To quantify the goodness of fit between the predicted values from theoretical models and measurements from cosmological observations, we use the $\chi^2$ statistic. By minimizing the $\chi^2$ function, one can identify the model parameters that best describe the observed universe.

For independent data points, the $\chi^2$ function is defined as
\begin{equation}
    \chi^2_\xi = \frac{(\xi_{th} - \xi_{obs})^2}{\sigma_\xi ^2},
\end{equation}
where $\xi_{th}$ is the theoretically predicted value, $\xi_{obs}$ is the experimentally measured value, and $\sigma_\xi$ is the standard deviation. For correlated data points, the $\chi^2$ function is given by
\begin{equation}
    \chi^2 = \Delta \xi ^T \text{Cov}^{-1} \Delta \xi,
\end{equation}
where $\Delta \xi \equiv \xi_{th} - \xi_{obs} $, and Cov$^{-1}$ is a inverse covariance matrix that characterizes the errors in the data.

Since we use the BAO data, the CMB distance priors data, and the SN data, the total $\chi^2$ is
\begin{equation}
    \chi^2 = \chi^2_{\text{BAO}} + \chi^2_{\text{CMB}} + \chi^2_{\text{SN}}.
\end{equation}
Assuming the measurement errors be Gaussian, the likelihood function is given by
\begin{equation}
    \mathcal{L}\propto e^{-\chi^2 /2}.
\end{equation}

To give a statistical interpretation of the results, we consider Akaike information criterion (AIC) \citep{Akaike:1974vps} and Bayesian information criterion (BIC) \citep{schwarz1978estimating}. 
The AIC is defined as
\begin{equation}
    \text{AIC} = -2\ln \mathcal{L}_{\text{max}} + 2k,
\end{equation}
where $k$ is the number of parameters, $\mathcal{L}_{\text{max}} = \chi^2_{\text{min}}$. 
The BIC is given by
\begin{equation}
    \text{BIC} = -2\ln \mathcal{L}_{\text{max}} + k \ln N,
\end{equation}
where $N$ is the number of data points used in the fit. 
Since only the relative vales of AIC and BIC between models are useful, we consider $\Delta \text{AIC}$ and $\Delta \text{BIC}$, where $\Delta \text{AIC}=\text{AIC}_\text{model}-\text{AIC}_{\Lambda \text{CDM}}$ and $\Delta \text{BIC}=\text{BIC}_\text{model}-\text{BIC}_{\Lambda \text{CDM}}$.
Compared to the reference model, a model with $\Delta \text{AIC} <2$ is substantially supported, a model with $4< \Delta \text{AIC} <7$ is less supported, and a model with $\Delta \text{AIC}>10$ is essentially not supported.
A difference of 2 in $\Delta \text{BIC}$ is considered significant evidence against the model with higher BIC value, while a difference of 6 provides stronger evidence.

In this work, we sample from the dark energy parameter posterior distributions using the MCMC code $\mathbf{Cobaya}$\footnote{https://github.com/CobayaSampler/cobaya} \citep{Torrado:2020dgo} with the $\mathbf{mcmc}$ sampler \citep{Lewis:2013hha, Lewis:2002ah}. Convergence of an MCMC run is assessed using the Gelman-Rubin statistic \citep{Gelman:1992zz} with a tolerance of $|R-1| < 0.02$. 
We analyze the MCMC chains using $\mathbf{Getdist}$  \citep{Lewis:2019xzd} to visualize the contour plots for the resulting posterior distributions. 
Table \ref{tab:flatprior} presents the flat prior ranges on which the parameters are left to freely vary. 
We use the current observational data to constrain these parameters and obtain the mean values and the 1-2 $\sigma$ confidence level.

\begin{table}
\begin{center}
\renewcommand{\arraystretch}{1.2}
\begin{tabular}{lll}
\hline \hline
$\textbf{Model}$ & $\textbf{Parameters}$ & $\textbf{Priors}$ \\
\hline \hline
{$\Lambda$CDM}
  &  $\Omega_m$      & $\mathcal{U}[0.01, 0.99]$ \\
  &  $\Omega_{b}h^2$ & $\mathcal{U}[0.005, 0.1]$ \\
  &  $h$         & $\mathcal{U}[0.2, 1.0]$ \\
\hline
{OHDE}
  &  $C$         & $\mathcal{U}[0.01, 2]$ \\
\hline
{GRDE}
  &  $\lambda$   & $\mathcal{U}[0.01, 2]$ \\
  &  $\beta$     & $\mathcal{U}[0.01, 2]$ \\
\hline
{IHDE1}
  &  $\epsilon$  & $\mathcal{U}[0.01, 2]$ \\
\hline
{IHDE2}
  &  $b^2$       & $\mathcal{U}[-1.0, 1.0]$ \\
  &  $C$         & $\mathcal{U}[0.01, 2]$ \\
\hline
{THDE}
  &  $\delta$    & $\mathcal{U}[1.01, 3]$ \\
\hline
{BHDE}
  &  $\Delta$    & $\mathcal{U}[0.01, 2]$ \\
\hline
\end{tabular}
\end{center}
\caption{Parameters and priors used in the analysis. $\mathcal{U}[a, b]$ represents an uniform distribution from a to b.}
\label{tab:flatprior}
\end{table}

\section{Holographic dark energy models and cosmology-fits}
\label{sec:HDE}
In this section, we briefly review the theoretical foundations of the six representative HDE models. In addition, we also present the corresponding cosmological fit results for each model. In this paper, we consider two primary data combinations: DESI+CMB (combining BAO and $Planck$ CMB data) and DESI+CMB+DESY5 (adding SN data from the DESY5 compilation).




\subsection{HDE models with other characteristic length scale}
This type of HDE model chooses the other characteristic length scale, which has nothing to do with the Hubble scale, as the IR cutoff.

\subsubsection{original holographic dark energy (OHDE) model}

In the OHDE model \citep{Li:2004rb}, an accelerating expanding universe is achieved by choosing the future event horizon as the characteristic length, defined as
\begin{equation}
    L = a\int^\infty_t \frac{d t'}{a} = a \int^\infty_a \frac{d a'}{H a'^2},
\end{equation}
where $a$ is the scale factor $a=(1+z)^{-1}$.
In this case, the Friedmann equation reads
\begin{equation}
\label{Fdmeq}
    3M_P^2 H^2 = \rho_r +\rho_{m} +\rho_{de},
\end{equation}
or equivalently,
\begin{equation}
\label{ez0}
    E(z) \equiv \frac{H(z)}{H_0} = \sqrt{ \frac{\Omega_r (1+z)^4 +\Omega_{m}(1+z)^3 }{1-\Omega_{de}(z)} },
\end{equation}
where $\Omega_r$ and $\Omega_{de}$ denote the fractional energy densities of radiation and dark energy, respectively. Taking derivative of $\Omega_{de}$, the dynamical evolution equation of $\Omega_{de}(z)$ is obtained as
\begin{equation}
\label{domdedz}
\begin{split}
    \frac{d \Omega_{de}(z)}{d z} = -\frac{2\Omega_{de}(z)(1-\Omega_{de}(z))}{1+z} \Big( & \frac{1}{2} + \frac{\sqrt{\Omega_{de}(z)}}{C} \\
    &+\frac{\Omega_{r}(z)}{2(1- \Omega_{de}(z))} \Big) .
\end{split}
\end{equation}
Solving Eq. (\ref{domdedz}) numerically and substituting the corresponding results into Eq. (\ref{ez0}), one can obtain the redshift evolution of $H(z)$ of the OHDE model, enabling the cosmological constraints to be obtained.

With conservation equation
\begin{equation}
\label{conserv.eq}
    \dot{\rho}_{de}+3H\rho_{de}(1+w)=0,
\end{equation}
and taking derivative for Eq. (\ref{rhode}) with respect to $x\equiv\ln a$, the EoS is given by
\begin{equation}
    w=-\frac{1}{3} - \frac{2\sqrt{\Omega_{de}}}{3C}.
\end{equation}
It is obvious that the EoS of the OHDE evolves dynamically and satisfies $-(1+2/C)/3 \leq w \leq -1/3$ due to $0 \leq \Omega_{de} \leq 1$.
In the late-time universe with $\Omega_{de} \simeq 1$, 
if $C= 1$, $w = -1$, then HDE will be close to the cosmological constant; 
if $C > 1$, $w > -1$, then HDE will be a quintessence DE; 
if $C < 1$, $w < -1$ thus HDE will be a phantom DE.
For more discussions on this model, we refer the reader to Ref. \citep{Wang:2016och, Wang:2023gov}.

\begin{figure}
    \centering
    \includegraphics[width=\linewidth]{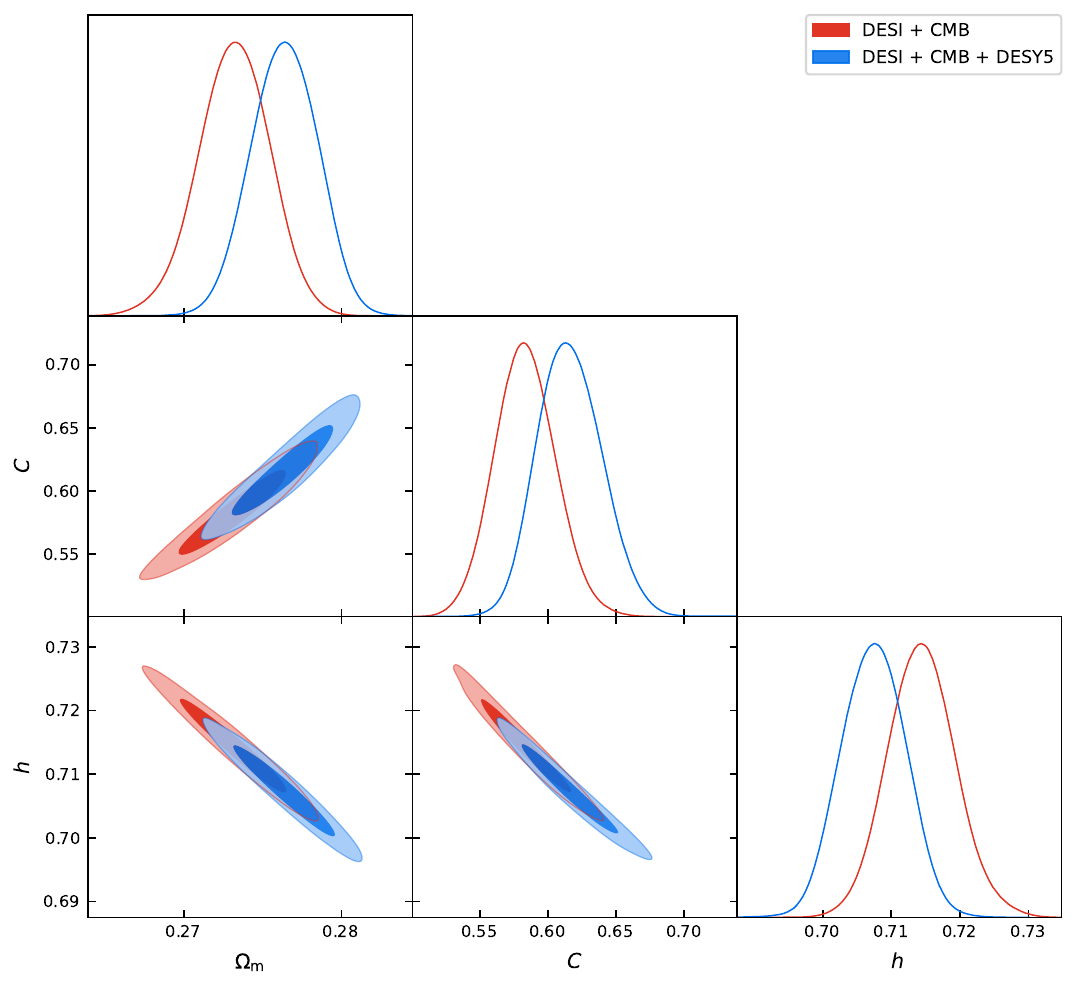}
    \caption{One-dimensional posterior distributions and two-dimensional marginalized contours at 1$\sigma$ and 2$\sigma$ levels for OHDE model.}
    \label{fig.OHDE}
\end{figure}

Figure \ref{fig.OHDE} presents the constraint result for the OHDE model. 
For the DESI+CMB dataset (red), the mean values are $ \Omega_m=0.2731 \pm 0.0022$, $ C=0.5833 \pm 0.0222$, $h =0.7144 \pm 0.0049$, which correspond to an EoS $w=-1.3077 \pm 0.0371$ and reduce the Hubble tension to $1.74 \sigma$. For the DESI+CMB+DESY5 dataset (blue), we have $ \Omega_m=0.2763 \pm 0.0020$, $C =0.6163 \pm 0.0235$, $ h=0.7074 \pm 0.0046$, corresponding to $w=-1.2535 \pm 0.0351$ and a Hubble tension of $2.49 \sigma$.


\subsection{HDE models with extended Hubble scale}

In these HDE models, IR cutoff is identified with a combination of Hubble scale and it's time derivative.

\subsubsection{Generalized Ricci dark energy (GRDE) model}
\cite{Granda:2008dk} proposed a generalization of the Ricci scalar as an IR cutoff, which is commonly known as the Granda-Oliver (GO) cutoff. In their model, the cutoff is expressed as
\begin{equation}
\label{gen-ricci}
    L^{-2} = \lambda H^2 + \beta \dot H,
\end{equation}
where $\lambda$ and $\beta$ are independent model parameters. Since the model takes a generalized Ricci scalar as the cutoff, we refer to it as the GRDE model. The corresponding HDE density is given by
\begin{equation}
\label{eq.rho_GRDE}
    \rho_{de} = 3 M_P^2(\lambda H^2 +\beta \dot H).
\end{equation}
By substituting Eq. (\ref{eq.rho_GRDE}) into the Friedmann equation and solving the resulting expression, the scaled Hubble parameter is given by
\begin{equation}
\begin{split}
    E(z)^2 = & \Omega_r (1+z)^4 + \Omega_m (1+z)^3 + \frac{2\beta - \lambda}{\lambda - 2\beta - 1} \Omega_r (1+z)^4 \\
    &+ \frac{3\beta -2\lambda}{2\lambda - 3\beta - 2} \Omega_m (1+z)^3 + g_0 (1+z)^{2(\lambda -1)/ \beta},
\end{split}
\end{equation}
where the last three terms give the scale dark energy density $\Omega_{de}$, $g_0$ can be determined by initial condition $E_0=E(t_0)=1$ as
\begin{equation}
    g_0 = 1 - \frac{2\Omega_m}{3\beta -2 \alpha +2} - \frac{\Omega_r}{2\beta -\alpha +1 }.
\end{equation}

With conservation equation, the EoS is given by
\begin{equation}
    w=\frac{2(\lambda-1)+(2-3\beta)\Omega_m +(2-4\beta)\Omega_r}{3\beta (1-\Omega_m -\Omega_r)}-1.
\end{equation}
Neglecting the radiation term, the GRDE model reduces to a cosmological constant when $\lambda =1$, $\beta = 2/3$. In the late-time universe, DE becomes the dominant component. Therefore, neglecting matter and radiation, if $0<\lambda -1< \beta$, the GRDE will be a quintessence DE; if $\lambda >1$ with $\beta <0$ or $\lambda <1$ with $\beta >0$, the GRDE will be a phantom DE. For more discussions on this model, we refer the reader to Ref. \citep{Granda:2008dk}.

\begin{figure}
    \centering
    \includegraphics[width=\linewidth]{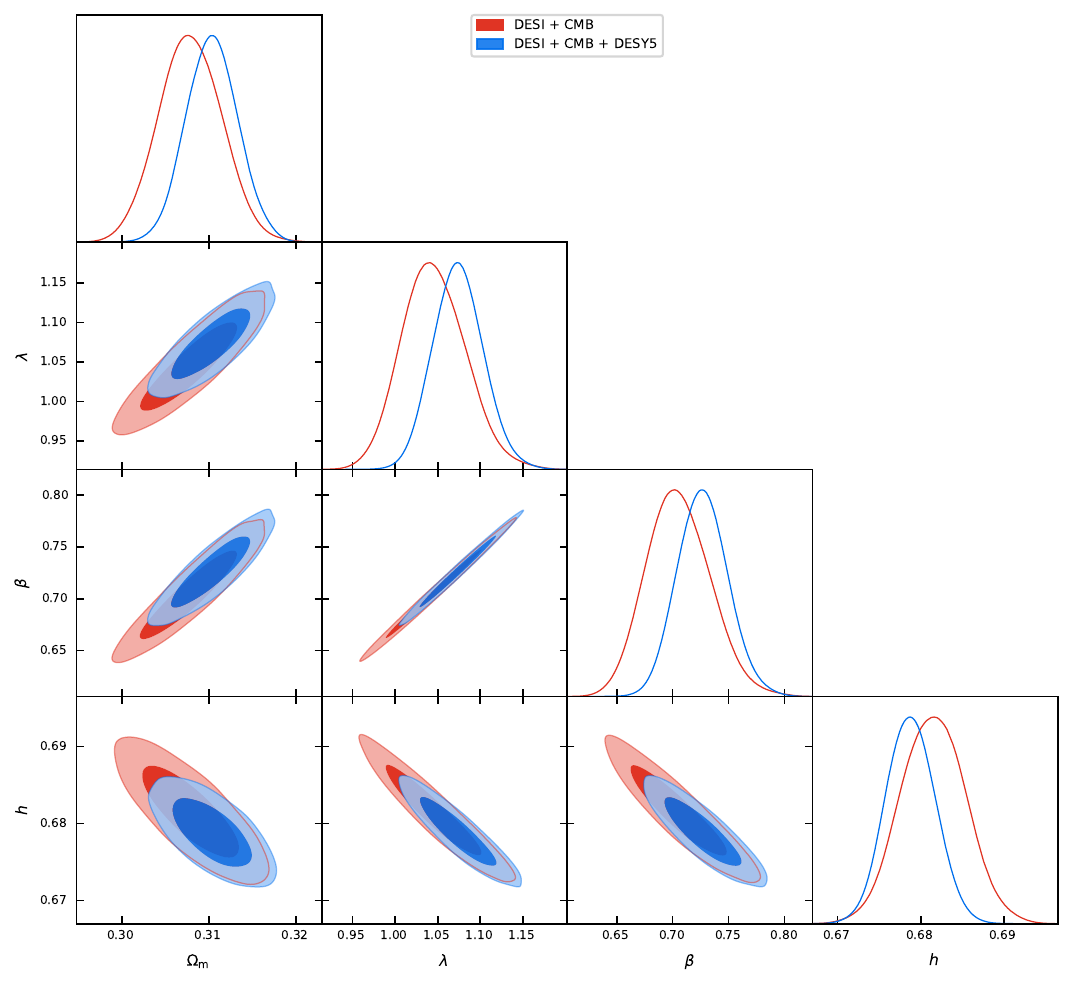}
    \caption{One-dimensional posterior distributions and two-dimensional marginalized contours at 1$\sigma$ and 2$\sigma$ levels for GRDE model}
    \label{fig.GRDE}
\end{figure}

Figure \ref{fig.GRDE} presents the constraint result for the GRDE model. For the DESI+CMB dataset (red), the mean values are $\Omega_m =0.3078 \pm 0.0036$, $\lambda =1.0455 \pm 0.0374$, $\beta =0.7052 \pm 0.0284$, $h =0.6815 \pm 0.0038$, which correspond to an EoS $w=-0.9622 \pm 0.0547$ and a Hubble tension of $5.33 \sigma$. For the DESI+CMB+DESY5 dataset (blue), we have $\Omega_m =0.3103 \pm 0.0029$, $\lambda =1.0741 \pm 0.0294$, $ \beta=0.7270 \pm 0.0225$, $h =0.6787 \pm 0.0029$, corresponding to $w=-0.9389 \pm 0.0422$ and a Hubble tension of $5.63 \sigma$.

\subsection{HDE models with dark sector interaction}
In these HDE models, dark energy and dark matter no longer evolve independently but interact with each other. We focus on two cases. In the first case, the Hubble horizon is adopted as the IR cutoff, referred to as the IHDE1 model. In the second case, the future event horizon is used as the IR cutoff, referred to as IHDE2 model.

\subsubsection{Interacting holographic dark energy (IHDE1) model with Hubble horizon}
Taking into account the mutual interaction, the energy densities of dark matter and DE evolve according to the equations below:
\begin{equation}
\label{Q1}
    \dot \rho_{dm} + 3H\rho_{dm}=Q,
\end{equation}
\begin{equation}
    \label{Q2}
    \dot \rho_{de} + 3H(1+w)\rho_{de}=-Q,
\end{equation}
where $Q$ denotes the interaction term. As shown in Ref. \citep{Pavon:2005yx}, the presence of interaction $Q$ enables the Hubble scale to serve as the cutoff length.

By adopting the Hubble scale as the IR cutoff and following the growth assumption in Ref. \citep{Zimdahl:2007zz}, the interaction term is given by
\begin{equation}
\label{IHDE1-Q}
    Q= 3\eta \rho_{m} H a ^\varepsilon,
\end{equation}
where $\varepsilon$ and $\eta$ are positive-definite parameters. Using the ansatz (\ref{IHDE1-Q}), the scaled Hubble rate is expressed as \citep{Zimdahl:2007zz}
\begin{equation}
\label{IHDE1-Hz}
    \frac{H(z)}{H_0} =  (1+z)^{3/2} \exp \left[ \frac{3\eta}{2\varepsilon} ( (1+z)^{-\varepsilon} -1) \right].
\end{equation}
Comparing Eq. (\ref{IHDE1-Hz}) with the corresponding quantity of the $\Lambda$CDM:
\begin{equation}
\label{ihde1-lcdm}
\begin{split}
    \frac{H(z)_{\Lambda CDM}}{H_0} &= \sqrt{\frac{\Omega_\Lambda}{\Omega_\Lambda + \Omega_m}} \left[ 1+\frac{\Omega_m}{\Omega_\Lambda} (1+z)^3 \right]^{1/2} \\
    &= 1 + \frac{3}{2}\frac{\Omega_m}{\Omega_\Lambda + \Omega_m}z + \mathcal{O}(z^2),
\end{split}
\end{equation}
we obtain the scaled Hubble rate for this IHDE1 model as follows
\begin{equation}
\label{ihde1-hz2}
\begin{split}
    E(z) &= (1+z)^{3/2} \exp \left[ \frac{3\eta }{2\varepsilon } ( (1+z)^{-\varepsilon} -1) \right] \\
    & = 1+ \frac{3}{2} (1- \eta) z + \mathcal{O}(z^2).
\end{split}
\end{equation}
As shown in Ref. \citep{Zimdahl:2007zz}, Eq. (\ref{ihde1-hz2}) must coincide with Eq. (\ref{ihde1-lcdm}) up to linear order in $z$. This requirement allows the parameter $\eta$ to be determined as
\begin{equation}
    \eta = \frac{\Omega_{de}}{\Omega_{de}+\Omega_m}.
\end{equation}
Once $\eta$ is determined, the scaled Hubble rate can be written as
\begin{equation}
    E(z) = (1+z)^{3/2} \exp \left[ \frac{3(1-\Omega_m) }{2\varepsilon } ( (1+z)^{-\varepsilon} -1) \right].
\end{equation}
In this model, the EoS is given by
\begin{equation}
    w = -\eta a^\varepsilon.
\end{equation}
Notice that DE is negligible in the early universe, it means that $\varepsilon >0$ must be satisfied. For the case of $\varepsilon >0$, the IHDE1 is a quintessence DE at present day. In the late-time universe with $\Omega_{de} \simeq 1$, the IHDE1 behaves as a cosmological constant. For more discussions on this model, we refer the reader to Ref. \citep{Zimdahl:2007zz}.

\begin{figure}
    \centering
    \includegraphics[width=\linewidth]{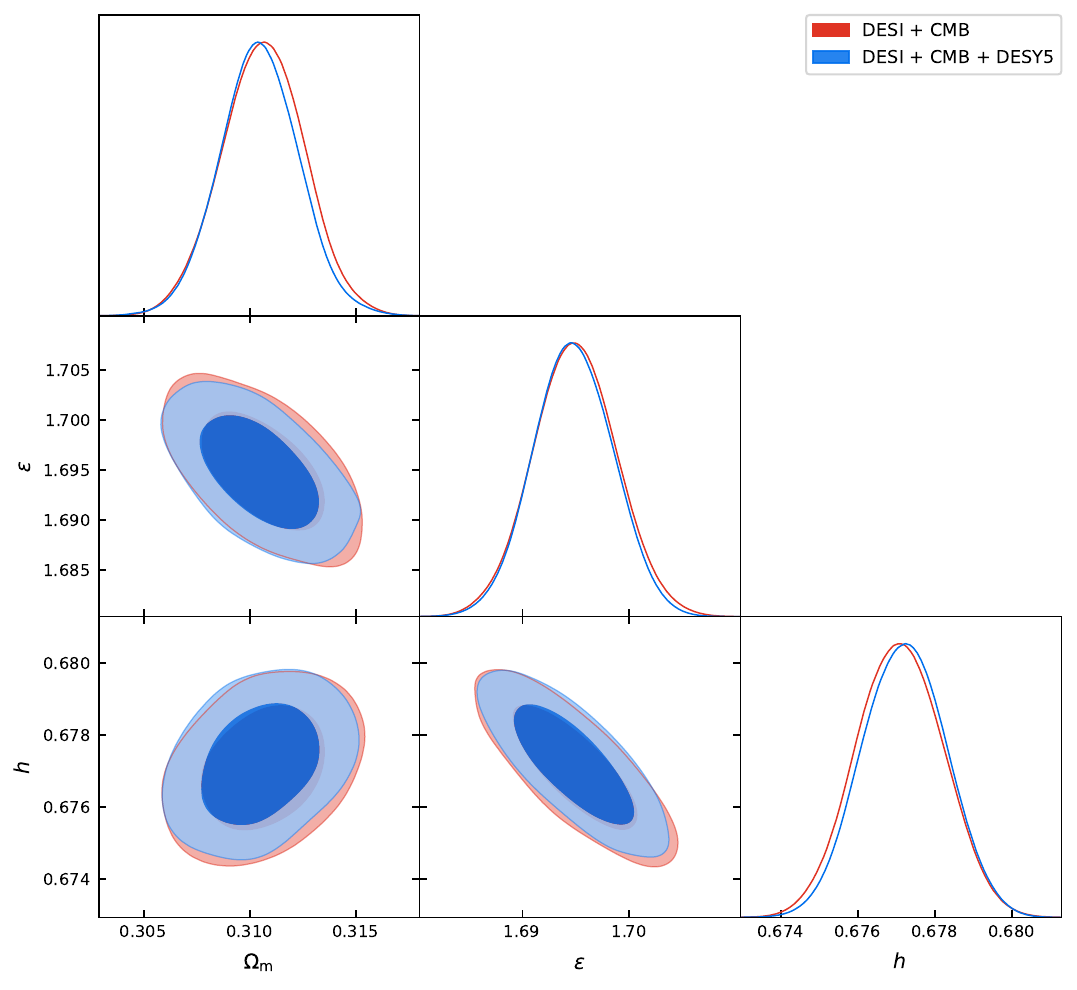}
    \caption{ One-dimensional posterior distributions and two-dimensional marginalized contours at 1$\sigma$ and 2$\sigma$ levels for IHDE1 model}
    \label{fig.IHDE1}
\end{figure}

Figure \ref{fig.IHDE1} presents the constraint result for the IHDE1 model. For the DESI+CMB dataset (red), the mean values are $\Omega_m =0.3106 \pm 0.0019$, $\epsilon =1.6949 \pm 0.0038$, $h =0.6771 \pm 0.0011$, which correspond to an EoS $w=-0.6894 \pm 0.0019$ and a Hubble tension of $6.29 \sigma$. For the DESI+CMB+DESY5 dataset (blue), the mean values are $\Omega_m =0.3104 \pm 0.0018$, $\epsilon =1.6947 \pm 0.0036$, $h =0.6770 \pm 0.0010$, corresponding to $w=-0.6896 \pm 0.0018$ and a Hubble tension of $6.31 \sigma$.

\subsubsection{Interacting holographic dark energy (IHDE2) model with future event horizon}
An alternative approach to the Hubble scale is to use the future event horizon as the IR cutoff. In this model, the interaction term takes the form $Q=3b^2 M_P^2 H (\rho_{de} +\rho_m)$ with $b^2$ the coupling constant \citep{Wang:2005jx}. By combining the interaction term and energy conservation equations (\ref{Q1}, \ref{Q2}), the evolution equation for $\Omega_{de}(z)$ is written as
\begin{equation}
    \frac{d \Omega_{de}}{d z}= -\frac{\Omega_{de}^2}{1+z} (1-\Omega_{de})\left[ \frac{1}{\Omega_{de}} +\frac{2}{C\sqrt{\Omega_{de}}} -\frac{3b^2 - \Omega_{r}}{\Omega_{de}(1-\Omega_{de})} \right].
\end{equation}
The Friedmann equation in this model satisfies
\begin{equation}
    \frac{1}{E(z)} \frac{dE(z)}{dz}= -\frac{\Omega_{de}}{1+z} (\frac{1}{2} +\frac{\sqrt{\Omega_{de}}}{C} +\frac{3b^2 -3-\Omega_r}{2\Omega_{de}} ).
\end{equation}
By numerically solving the above equation, we then obtain the evolutions of both $\Omega_{de}$ and $E(z)$ as functions of redshift.

The EoS is given by
\begin{equation}
    w=-\frac{1}{3} - \frac{2\sqrt{\Omega_{de}}}{3C} -\frac{b^2}{\Omega_{de}}.
\end{equation}
If $b^2 <0$, it means a transfer of energy from dark matter to DE,
if $b^2 >0$, it means a transfer of energy from DE to dark matter,
if $b^2 =0$, the EoS reduces to the OHDE model. In the late-time universe with $\Omega_{de} \simeq 1$, if we neglect the interaction, the behaviors of IHDE2 will be the same as OHDE. For more discussions on this model, we refer the reader to Ref. \citep{Wang:2005jx}.

\begin{figure}
    \centering
    \includegraphics[width=\linewidth]{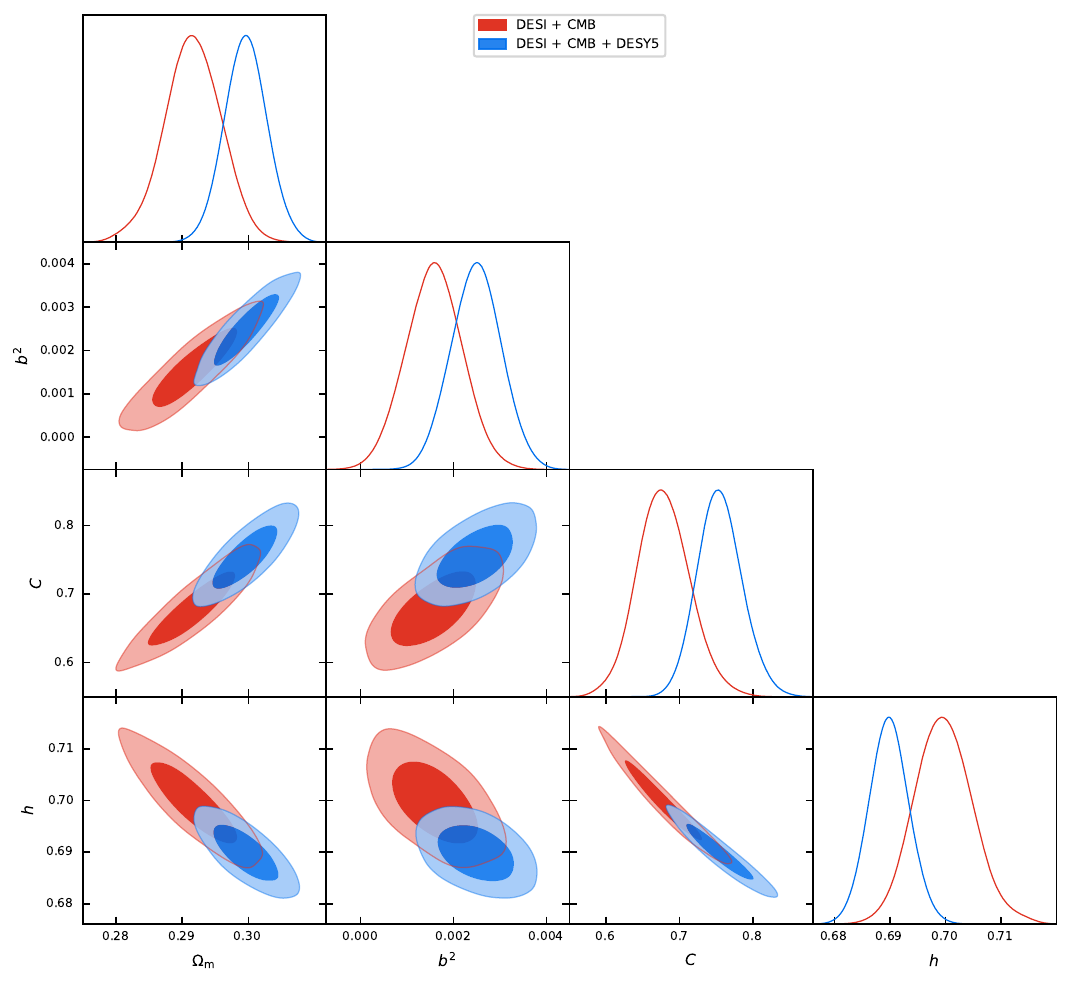}
    \caption{ One-dimensional posterior distributions and two-dimensional marginalized contours at 1$\sigma$ and 2$\sigma$ levels for IHDE2 model}
    \label{fig.IHDE2}
\end{figure}

Figure \ref{fig.IHDE2} presents the constraint result for the IHDE2 model. For the DESI+CMB dataset (red), the mean values are $\Omega_m =0.2915 \pm 0.0043$, $b^2 =0.0015 \pm 0.0006$, $C =0.6784 \pm 0.0361$, $h =0.6996 \pm 0.0052$, which correspond to an EoS $w=-1.1625 \pm 0.0441$ and reduce the Hubble tension to $3.19 \sigma$. For the DESI+CMB+DESY5 dataset (blue), the mean values are $\Omega_m =0.2996 \pm 0.0032$, $b^2 =0.0024 \pm 0.0005$, $C =0.7548 \pm 0.0305$, $h =0.6898 \pm 0.0036$, corresponding to $w=-1.0759 \pm 0.0299$ and a Hubble tension of $4.49 \sigma$.

\subsection{HDE models with modified black hole entropy}

In these HDE models, the HDE density depends on the specific entropy-area relationship $S \sim A \sim L^2$ of black holes, where $A=4\pi L^2$ represents the area of the horizon.

\subsubsection{Tsallis holographic dark energy (THDE) model with Hubble horizon}
Tsallis and Cirto proposed that the traditional Boltzmann-Gibbs additive entropy should be generalized to a non-additive entropy, known as Tsallis entropy \citep{Tsallis:2012js}. This generalized entropy is given by
\begin{equation}
\label{S_Tsallis}
    S_\delta =\gamma A^\delta,
\end{equation}
where $\gamma$ is an unknown constant, and $\delta$ denotes the non-additivity parameter. Building on this concept, M. Tavayef et al. introduced the Tsallis HDE (THDE) model \citep{Tavayef:2018xwx}. By substituting the relation (\ref{S_Tsallis}) into the ultraviolet (UV) and IR (L) cutoff $ \rho_{de} \leq S L^{-4}$, the energy density of DE is modified as
\begin{equation}
    \rho_{de} = B L^{2\delta -4},
\end{equation}
where $B = 3C^2M_P^2$. In this model, the Hubble scale is considered a suitable candidate for the IR cutoff, which can lead to the late-time accelerated expansion of the universe. The Friedmann equation of the THDE model also satisfies
\begin{equation}
\label{Fmeq}
    E(z)= \sqrt{ \frac{ \Omega_r(1+z)^4 + \Omega_{m}(1+z)^3 }{1-\Omega_{de}(z)} }.
\end{equation}
The evolution equation of $\Omega_{de}$ is governed by the equation
\begin{equation}
    \frac{d \Omega_{de}}{d z}= - \frac{3(\delta -1)}{1+z} \Omega_{de} \left( \frac{1-\Omega_{de} - 5\Omega_{r}}{1-(2-\delta)\Omega_{de}} \right).
\end{equation}

With conservation equation (\ref{conserv.eq}), the EoS is given by
\begin{equation}
    w=\frac{\delta-1}{(2-\delta)\Omega_{de}-1}.
\end{equation}
When $\delta =2$, the EoS of this model coincides with that of $\Lambda$CDM, where $w=-1$. 
Assuming the present-day $\Omega_{de}=0.7$, 
if $\delta >2$, the THDE will be a quintessence DE,
if $\delta <2$, the THDE will be a phantom DE.
For more discussions on this model, we refer the reader to Ref. \citep{Tavayef:2018xwx}.

\begin{figure}
    \centering
    \includegraphics[width=\linewidth]{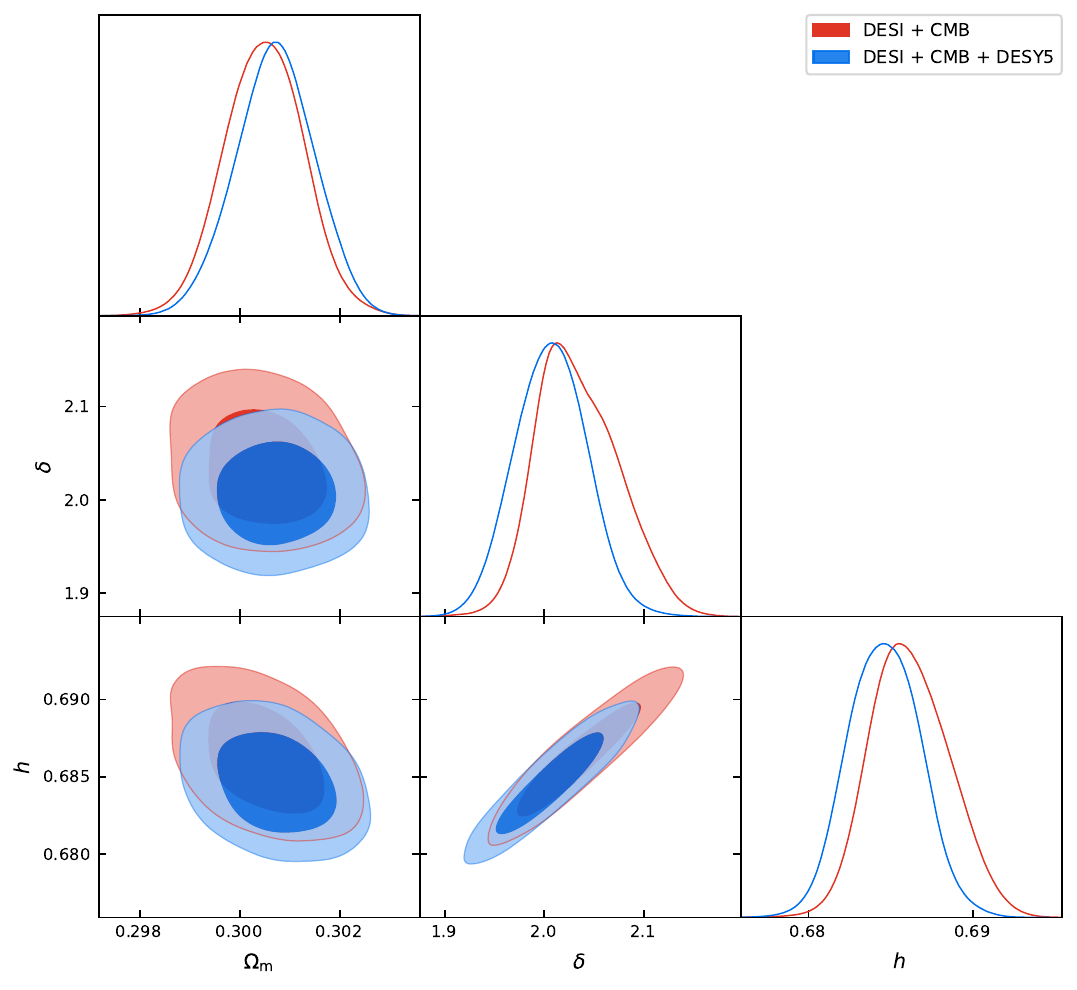}
    \caption{ One-dimensional posterior distributions and two-dimensional marginalized contours at 1$\sigma$ and 2$\sigma$ levels for THDE model}
    \label{fig.THDE}
\end{figure}

Figure \ref{fig.THDE} presents the constraint result for the THDE model. For the DESI+CMB dataset (red), the mean values are $\Omega_m =0.3004 \pm 0.0008$, $\delta =2.0344 \pm 0.0409$, $h =0.6861 \pm 0.0023$, which correspond to an EoS $w=-1.0101 \pm 0.0117$ and a Hubble tension of $5.12 \sigma$. For the DESI+CMB+DESY5 dataset (blue), the mean values are $\Omega_m =0.3007 \pm 0.0007$, $\delta =2.0068 \pm 0.0361$, $h =0.6846 \pm 0.0021$, corresponding to $w=-1.0020 \pm 0.0108$ and a Hubble tension of $5.32 \sigma$.

\subsubsection{Barrow holographic dark energy (BHDE) model with future event horizon}
In Refs. \citep{Barrow:2020tzx, Barrow:2020kug}, it was suggested that quantum-gravitational effects may cause deformations on the black hole surface, leading to deviations from the standard Bekenstein-Hawking entropy. This modified entropy, known as Barrow entropy, is expressed as
\begin{equation}
    S_B=\left( \frac{A}{A_0} \right)^{1+\Delta/2},
\end{equation}
where $A$ is the standard horizon area, $A_0$ is the Planck area, and $\Delta$ is a parameter quantifies quantum-gravitational deformation. Here, $0 \leq \Delta \leq 1$, with $\Delta = 0$ corresponding to a smooth spacetime structure and $\Delta = 1$ representing the most intricate deformation. By applying Barrow entropy to UV/IR relation, \cite{Saridakis:2020zol} proposed the Barrow HDE (BHDE) model. In this model, the HDE density is modified as
\begin{equation}
    \rho_{de}=B L^{\Delta-2}.
\end{equation}
We fix $B=3$ in $M_P^2$ units to perform constraints. In the BHDE model, the future event horizon serves as the IR cutoff. The Friedmann equation also satisfies Eq. (\ref{Fmeq}), and the evolution equation for $\Omega_{de}$ is \citep{Denkiewicz:2023hyj}
\begin{equation}
\begin{split}
    \frac{d \Omega_{de}}{d z} = &-\frac{\Omega_{de}(1-\Omega_{de})}{1+z} [ (1+\frac{\Delta}{2})\mathcal{F}_r + (\Delta +1)\mathcal{F}_m \\
    &+G(1-\Omega_{de})^{\frac{\Delta}{2(\Delta -2)}} (\Omega_{de})^{\frac{1}{2-\Delta}} ],
\end{split}
\end{equation}
where
\begin{equation}
\begin{split}
    \mathcal{F}_r &=\frac{2\Omega_r (1+z)^4}{\Omega_m (1+z)^3 + \Omega_r (1+z)^4},\\
    \mathcal{F}_m &=\frac{\Omega_m (1+z)^3}{\Omega_m (1+z)^3 + \Omega_r (1+z)^4},\\
    G &\equiv (2-\Delta) ( \frac{B}{3M_P^2} )^{\frac{1}{\Delta -2}} ( H_0 \sqrt{\Omega_m (1+z)^3 + \Omega_r (1+z)^4 })^{\frac{\Delta}{2-\Delta}}.
\end{split}
\end{equation}
The EoS is given by
\begin{equation}
    w=-\frac{1+\Delta}{3}-\frac{G}{3}(\Omega_{de})^{\frac{1}{2-\Delta}} (1-\Omega_{de})^{\frac{\Delta}{2(\Delta-2)}} e^{\frac{3\Delta}{2(2-\Delta)}x}.
\end{equation}
In the standard case of $\Delta =0$, the EoS reduces to that of the OHDE model. 
For $\Delta \lesssim 0.5$, the BHDE behaves as a quintessence DE,
for $\Delta >0.5$, the BHDE behaves as a quintom DE. 
In the late-time universe with $\Omega_{de} \simeq 1$, if $\Delta =0$, the behaviors of BHDE will be the same as OHDE.
For more discussions on this model, we refer the reader to Ref. \citep{Saridakis:2020zol}.

\begin{figure}
    \centering
    \includegraphics[width=\linewidth]{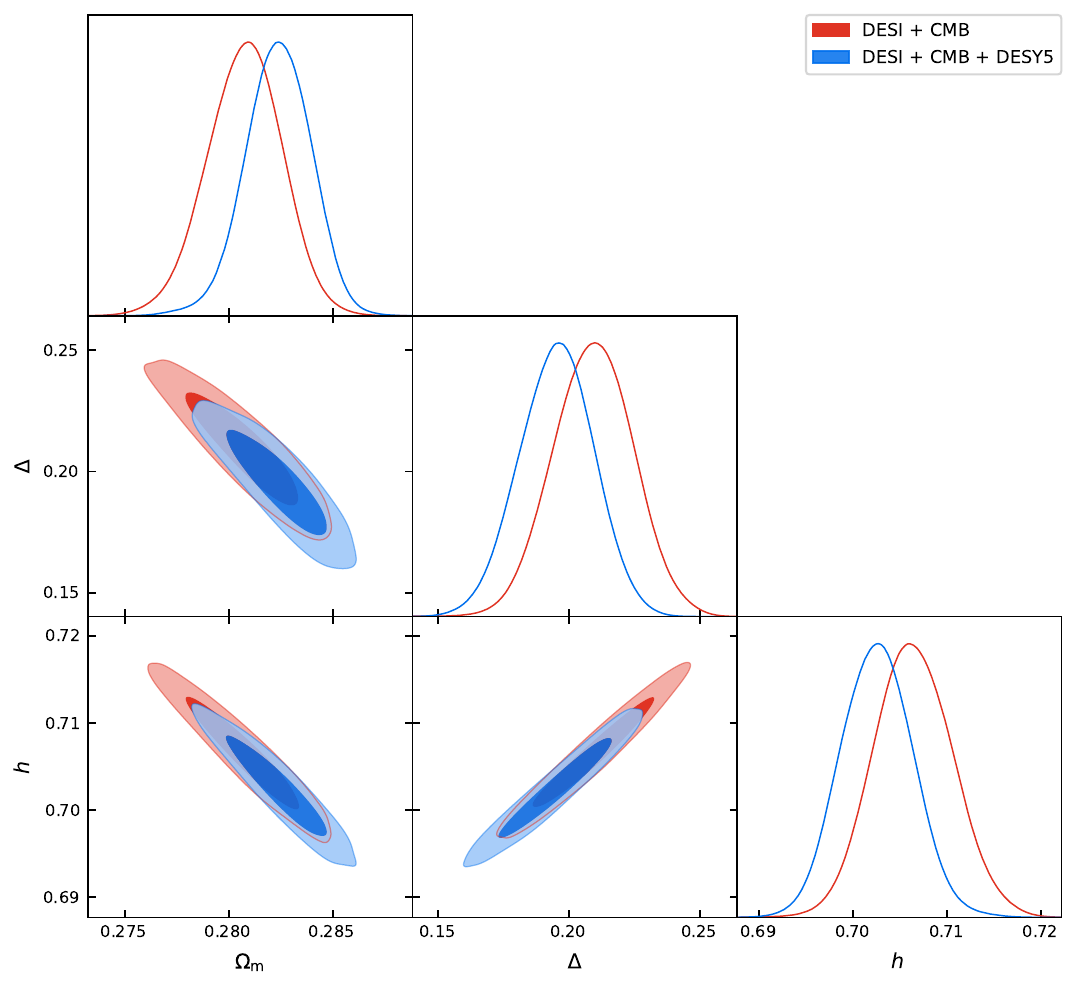}
    \caption{One-dimensional posterior distributions and two-dimensional marginalized contours at 1$\sigma$ and 2$\sigma$ levels for BHDE model.}
    \label{fig.BHDE}
\end{figure}

Figure \ref{fig.BHDE} presents the constraint result for the BHDE model. For the DESI+CMB dataset, the mean values are $\Omega_m =0.2807 \pm 0.0018$, $\Delta =0.2091 \pm 0.0152$, $h =0.7064 \pm 0.0041$, which correspond to an EoS $w=-1.2195 \pm 0.0298$ and reduce the Hubble tension to $2.65 \sigma$. For the DESI+CMB+DESY5 dataset, the mean values are $\Omega_m =0.2823 \pm 0.0015$, $\Delta =0.1951 \pm 0.0140$, $h =0.7025 \pm 0.0037$, corresponding to $w=-1.1911 \pm 0.0264$ and a Hubble tension of $3.12 \sigma$.

The THDE and BHDE models are closely related to generalized entropy constructions that also appear in modified gravity or non-extensive thermodynamic frameworks \citep{Hernandez-Almada:2021aiw, Basilakos:2023kvk, Basilakos:2023seo}.
In such frameworks, the choice of IR cutoff still plays an important role in HDE models.
Therefore, similar trends regarding the IR cutoff and its impact on the inferred value of $H_0$ are likely to persist.

\section{Discussions on the Hubble tension problem in the framework of HDE}
\label{sec:results}
In this section, based on the cosmological results of Section \ref{sec:HDE}, we discuss the Hubble tension problem in the framework of HDE. For comparison, we also present results for the $\Lambda$CDM model. 

For clarity, in this work we interpret the residual Hubble tension as a significant alleviation for $\Delta \sigma < 1 \sigma$, a partial mitigation for $1\sigma \leq \Delta \sigma < 3 \sigma$, a marginal improvement for $3\sigma \leq \Delta \sigma < 5 \sigma$, and no alleviation for $\Delta \sigma \geq 5\sigma$. The adopted criteria are summarized in Table \ref{tab:tension}.

\begin{table}
\begin{center}
\renewcommand{\arraystretch}{1.5}
\begin{tabular}{c|c}
\hline \hline
$\text{Residual Tension}$ & $\text{Degree of Alleviation}$ \\
\hline
$\Delta \sigma < 1 \sigma$ & significant alleviation \\
$1\sigma \leq \Delta \sigma < 3 \sigma$ & partial mitigation \\
$3\sigma \leq \Delta \sigma < 5 \sigma$ & marginal improvement \\
$\Delta \sigma \geq 5\sigma$ & no alleviation \\
\hline \hline
\end{tabular}
\end{center}
\caption{Criteria adopted in this work for interpreting the residual Hubble tension.}
\label{tab:tension}
\end{table}

\subsection{Reslts given by the DESI+CMB and the DESI+CMB+DESY5 datasets}
\begin{figure}
    \centering
    \includegraphics[width=0.5\textwidth]{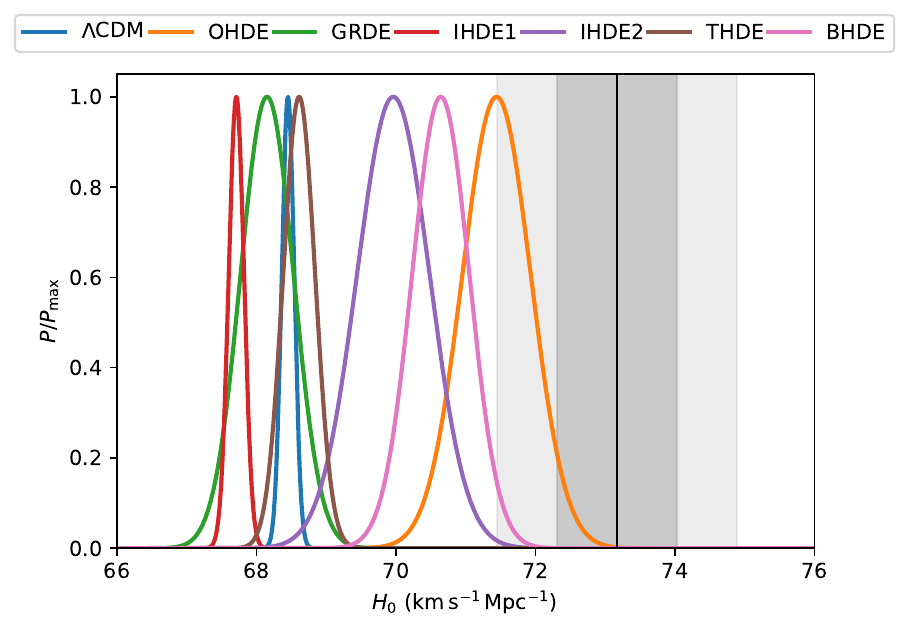}
    \caption{One-dimensional marginalized posterior distributions for $H_0$ across different DE models using $\textbf{DESI+CMB}$ dataset. The vertical gray line marks the SH0ES measurements of $H_0$ = 73.17 $\pm$ 0.86 $\text{km}$ $\text{s}^{-1}$ $\text{Mpc}^{-1}$. The shaded bands denote the $1\sigma$ (dark gray) and $2\sigma$ (light gray) CL around this value.}
    \label{fig.DESI+CMB}
\end{figure}

\begin{figure}
    \centering
    \includegraphics[width=0.5\textwidth]{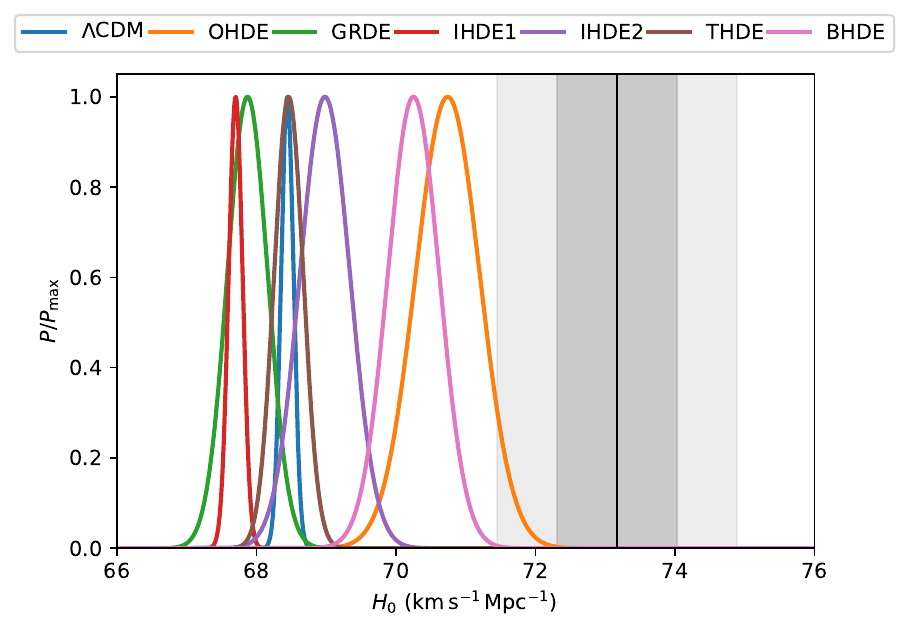}
    \caption{One-dimensional marginalized posterior distributions for $H_0$ across different DE models using $\textbf{DESI+CMB+DESY5}$ dataset.}
    \label{fig.DESI+CMB+D5}
\end{figure}

To illustrate the ability of different models of alleviating the Hubble tension problem, we present the one-dimensional marginalized posterior distributions of $H_0$ in Figs. \ref{fig.DESI+CMB} and \ref{fig.DESI+CMB+D5}.
Fig. \ref{fig.DESI+CMB} shows results for the DESI+CMB dataset, while Fig. \ref{fig.DESI+CMB+D5} shows results for the DESI+CMB+DESY5 dataset.
The curves represent different models: $\Lambda$CDM (blue), OHDE (orange), GRDE (green), IHDE1 (red), IHDE2 (purple), THDE (brown), and BHDE (pink).
The vertical gray line marks the SH0ES measurements of $H_0$ = 73.17 $\pm$ 0.86 $\text{km}$ $\text{s}^{-1}$ $\text{Mpc}^{-1}$, with shaded bands denoting the $1\sigma$ (dark gray) and $2\sigma$ (light gray) CL.
In addition, we summarize the corresponding marginalized cosmological parameter constraints (with 1$\sigma$ error) and the statistical significance of the $H_0$ tension with respective to the SH0ES measurement in Table \ref{tab:result_BC} and \ref{tab:result_BCS}. 
In the following, we focus on these two cases: DEIS+CMB and DESI+CMB+DESY5.

\begin{table}
    \centering
    \begin{tabular}{lccc}
    \hline \hline
      & \multicolumn{3}{c}{DESI + CMB}        \\
    \cline{2-4}
      \textbf{Model}   & $\Omega_m$  &  $H_0$  & $H_0$ tension \\
     \hline 
      $\Lambda$CDM    & 0.3004$\pm$0.0007   & 68.45$\pm$0.09   & 5.45$\sigma$ \\
      \hline 
      OHDE        & 0.2731$\pm$0.0022  & 71.44$\pm$0.49   & 1.74$\sigma$ \\
      BHDE       & 0.2807$\pm$0.0018  & 70.64$\pm$0.41   & 2.65$\sigma$ \\
      IHDE2       & 0.2915$\pm$0.0043  & 69.96$\pm$0.52   & 3.19$\sigma$ \\
      THDE       & 0.3004$\pm$0.0008  & 68.61$\pm$0.23   & 5.12$\sigma$ \\
      GRDE        & 0.3078$\pm$0.0036  & 68.15$\pm$0.38   & 5.33$\sigma$ \\
      IHDE1       & 0.3106$\pm$0.0019  & 67.71$\pm$0.11   & 6.29$\sigma$ \\
     \hline \hline \\
    \end{tabular}
    \\[+1mm]
    \begin{flushleft}
    \caption{Cosmological parameters constraints from the $\textbf{DESI+CMB}$ dataset for the baseline flat $\Lambda$CDM model and six representative HDE models. The $H_0$ tension significance is calculated with respect to the SH0ES measurement. For the HDE models, the statistical significance of tension is shown in increasing order.}
    \label{tab:result_BC}
    \end{flushleft}
\end{table}

From Fig. \ref{fig.DESI+CMB} and Table \ref{tab:result_BC}, one can find that the posterior mean of $H_0$ in the $\Lambda$CDM model is significantly lower than the SH0ES measurement, giving a tension of $5.45 \sigma$.
This reveals that there is a significant Hubble tension problem for the $\Lambda$CDM model.
Three HDE models \textemdash{} THDE, GRDE, and IHDE1 \textemdash{} give relatively low $H_0$ values that are close to the result of the $\Lambda$CDM model.
Therefore, they also have a significant Hubble tension problem with tensions of $5.12\sigma$, $5.33\sigma$, and $6.29\sigma$, respectively.
Notably, these three models have a common feature: they adopt the Hubble scale or its combinations as the IR cutoff, showing that models with this feature have difficulty resolving the tension.
In contrast, the OHDE, BHDE, and IHDE2 models have higher $H_0$ values that are much closer to the SH0ES value.
Therefore, the Hubble tensions are alleviated to $1.74\sigma$, $2.65\sigma$, and $3.19\sigma$, respectively.
These models employ the future event horizon as the IR cutoff, demonstrating that the choice of future event horizon can partially mitigate the Hubble tension.

After presenting results for the DESI+CMB combination in Fig. \ref{fig.DESI+CMB} and Table \ref{tab:result_BC}, we now examine the effects of incorporating SN data.
Figure \ref{fig.DESI+CMB+D5} and Table \ref{tab:result_BCS} show results for the DESI+CMB+DESY5 dataset. 
Compared to DESI+CMB, all the posterior distributions of $H_0$ shift slightly toward lower values, but the overall trend is similar.
For the $\Lambda$CDM model, the mean value of $H_0$ remains far from the SH0ES measurement, with a tension of $5.47\sigma$.
This reveals that for the case of DESI+CMB+DESY5, the Hubble tension problem is still significant for the $\Lambda$CDM model.
In addition, HDE models based on the Hubble scale or its combinations (THDE, GRDE, and IHDE1) continue to give relatively low $H_0$ values, with tensions of $5.32\sigma$, $5.84\sigma$, and $6.31\sigma$, respectively.
Therefore, these results reaffirm that HDE models based on the Hubble scale or its combinations cannot alleviate the tension.
In contrast, HDE models based on the future event horizon (OHDE, BHDE, and IHDE2) maintain higher $H_0$ values closer to the SH0ES measurement, alleviating tensions to $2.49\sigma$, $3.12\sigma$, and $4.49\sigma$, respectively. 
Comparing tensions between the two data combinations reveals that incorporating the DESY5 compilation increases the Hubble tension for all models.

\begin{table}
    \centering
    \renewcommand{\arraystretch}{1.0}
    \begin{tabular}{lccc|c}
    \hline \hline
      & \multicolumn{3}{c}{DESI+CMB+DESY5}  & \multicolumn{1}{|c}{DESI+CMB} \\
    \cline{2-4} \cline{5-5}
      \textbf{Model}   & $\Omega_m$  &  $H_0$  & $H_0$ tension  & $H_0$ tension \\
     \hline 
      $\Lambda$CDM    & 0.3006$\pm$0.0007  & 68.44$\pm$0.09   & 5.47$\sigma$  & 5.45$\sigma$ \\
      \hline 
      OHDE        & 0.2763$\pm$0.0020  & 70.74$\pm$0.46   & 2.49$\sigma$  & 1.74$\sigma$ \\
      BHDE        & 0.2823$\pm$0.0015  & 70.25$\pm$0.37   & 3.12$\sigma$  & 2.65$\sigma$ \\
      IHDE2       & 0.2996$\pm$0.0032  & 68.98$\pm$0.36   & 4.49$\sigma$  & 3.19$\sigma$ \\
      THDE        & 0.3007$\pm$0.0007  & 68.46$\pm$0.21   & 5.32$\sigma$  & 5.12$\sigma$ \\
      GRDE        & 0.3103$\pm$0.0029  & 67.87$\pm$0.29   & 5.84$\sigma$  & 5.33$\sigma$ \\
      IHDE1       & 0.3104$\pm$0.0018  & 67.70$\pm$0.10   & 6.31$\sigma$  & 6.29$\sigma$ \\
     \hline \hline \\[-1mm]
    \end{tabular}
    \\[+1mm]
    \begin{flushleft}
    \caption{Cosmological parameters constraints from the $\textbf{DESI+CMB+DESY5}$ dataset for the baseline flat $\Lambda$CDM model and six representative HDE models. For direct comparison, the tension derived from the DESI+CMB dataset is also provided in the last column.}
    \label{tab:result_BCS}
    \end{flushleft}
\end{table}

For further discussion, the results of $\Delta \text{AIC}$ and $\Delta \text{BIC}$ are listed in Table \ref{tab:AIC_BIC}.
Combining the results in Table \ref{tab:result_BCS} with those in Table \ref{tab:AIC_BIC}, one can see a common trend.
The HDE models that employ the Hubble scale or its combination have relatively better statistical performance (such as GRDE and THDE models); these models cannot alleviate the Hubble tension.
In contrast, the HDE models that employ the future event horizon show worse statistical performance (such as OHDE, BHDE, and IHDE2 models); these models can reduce the Hubble tension. 
The key difference is that the future event horizon cutoff is more complicated than the Hubble scale cutoff.
Therefore, for the OHDE, BHDE, and IHDE2 models, the improved agreement in $H_0$ is achieved at the expense of increased model complexity.

\begin{table}
    \centering
    \renewcommand{\arraystretch}{1.1}
    \begin{tabular}{lcccccc}
    \hline \hline
    & \multicolumn{3}{c}{DESI+CMB}  & \multicolumn{3}{|c}{DESI+CMB+DESY5} \\
    \cline{2-4} \cline{5-7}
      \textbf{Model}  & $\chi^2_{\text{min}}$ & $\Delta$AIC  & $\Delta$BIC  & $\chi^2_{\text{min}}$ & $\Delta$AIC  & $\Delta$BIC \\
     \hline 
      $\Lambda$CDM   & 16.22 & 0  & 0  & 1661.13 & 0  & 0 \\
      \hline
      GRDE        & 10.00 & -2.22  & -1.61   & 1649.22 & -7.91  & 3.12  \\
      THDE        & 15.38 & 1.16   & 1.46   & 1657.96 & -1.17  & 4.34  \\
      IHDE2       & 19.11 & 6.89   & 7.49   & 1665.60 & 8.47   & 19.5  \\
      BHDE        & 34.11 & 19.89  & 20.19  & 1681.27 & 22.14  & 27.65  \\
      OHDE        & 43.93 & 29.71  & 30.01  & 1696.19 & 37.06  & 42.57  \\
      IHDE1       & 55.94 & 41.72  & 42.02  & 1695.30 & 36.17  & 41.68  \\
     \hline \hline \\[-1mm]
    \end{tabular}
    \\[+1mm]
    \begin{flushleft}
    \caption{Summary of the information criteria results. The relative values $\Delta$AIC and $\Delta$BIC are computed with respect to the $\Lambda$CDM model. The DESI+CMB dataset contains 10 data points, while the DESI+CMB+DESY5 dataset contains 1839 data points.}
    \label{tab:AIC_BIC}
    \end{flushleft}
\end{table}

In summary, HDE models that employ the Hubble scale or its combinations as the IR cutoff cannot alleviate the Hubble tension problem.
In contrast, HDE models that employ the future event horizon as the IR cutoff can partially mitigate the Hubble tension.
Furthermore, incorporating the DESY5 compilation into the DESI+CMB dataset increases the tension.

\subsection{Effects of adopting alternative BAO and SN data}
\label{sec:discussion}
\begin{figure}
    \centering
    \includegraphics[width=0.5\linewidth]{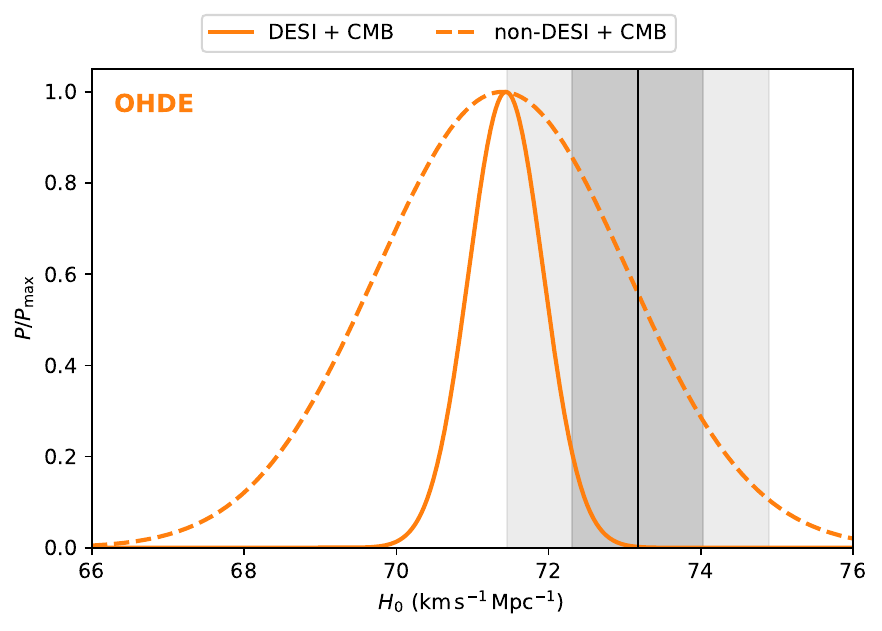}
    
    \includegraphics[width=0.45\linewidth]{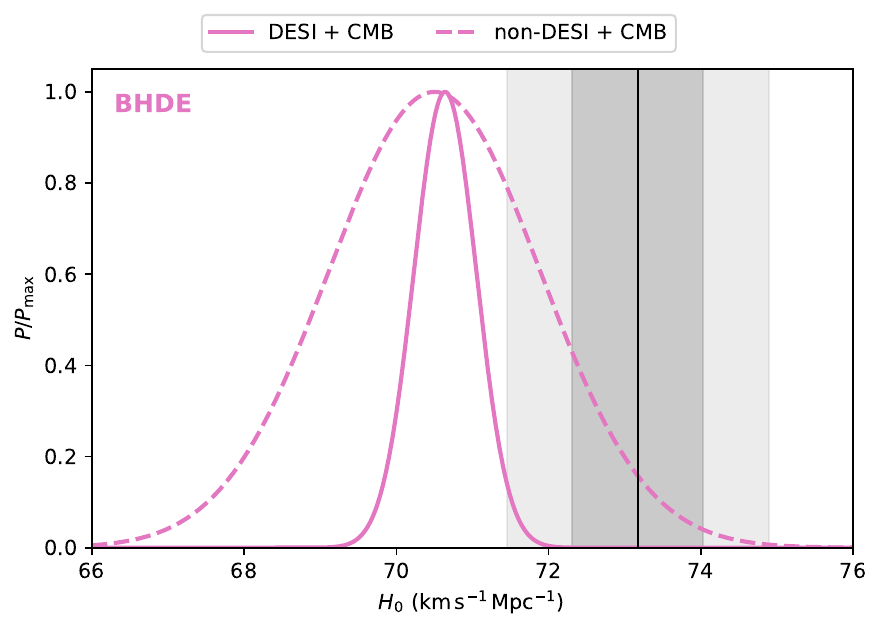}
    \includegraphics[width=0.45\linewidth]{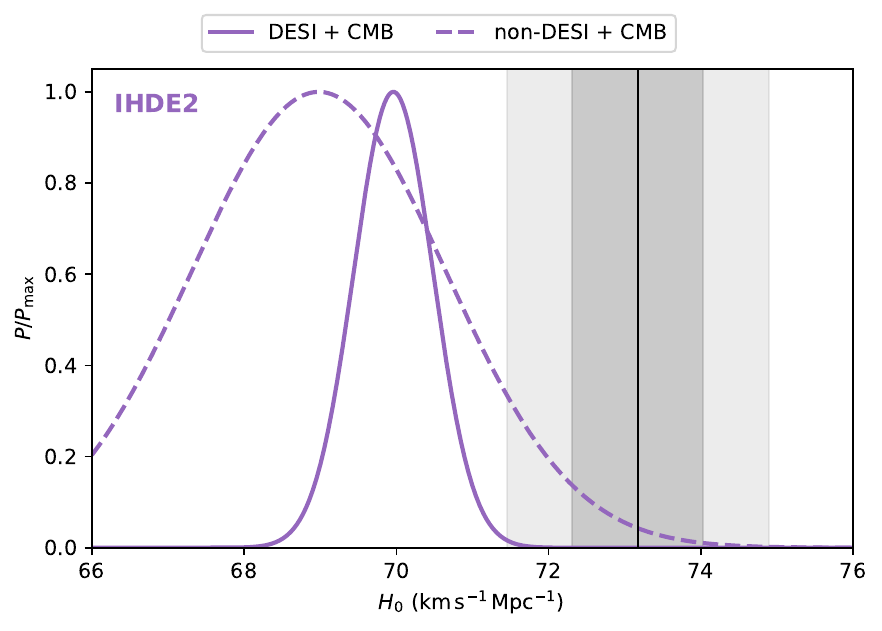}

    \caption{ One-dimensional marginalized posterior distributions for $H_0$ in the OHDE, BHDE, and IHDE2 models.  Solid curves denote the $\textbf{DESI+CMB}$ dataset, while dashed curves correspond to the $\textbf{non-DESI+CMB}$ dataset. }
    \label{fig.BAO_comparison}
\end{figure}

\begin{figure}
    \centering
        \includegraphics[width=0.5\linewidth]{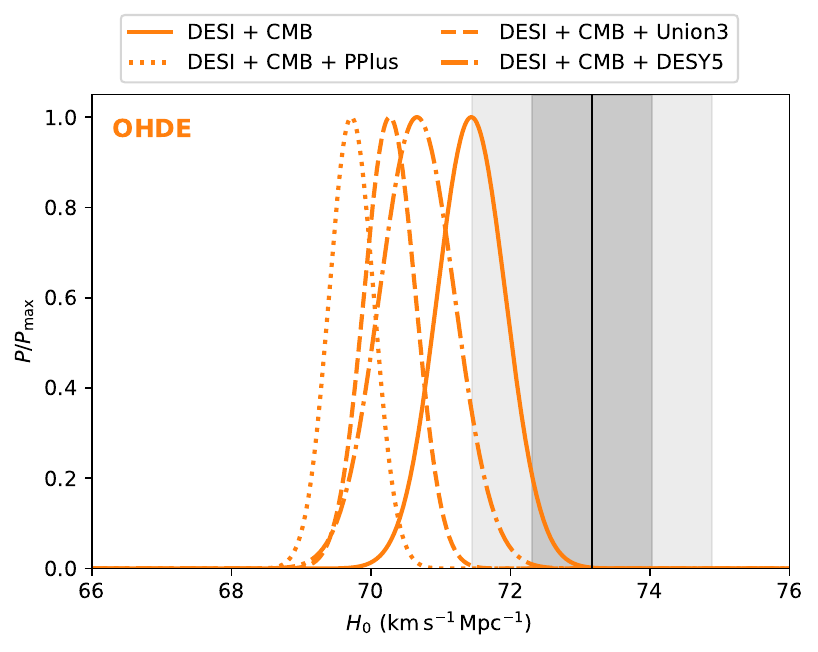}
    
        \includegraphics[width=0.45\linewidth]{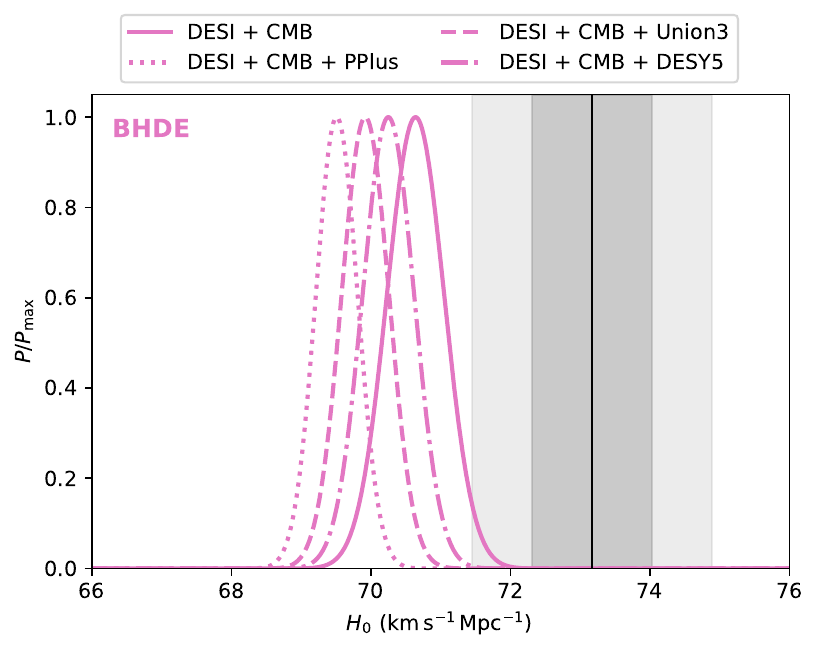}
        \includegraphics[width=0.45\linewidth]{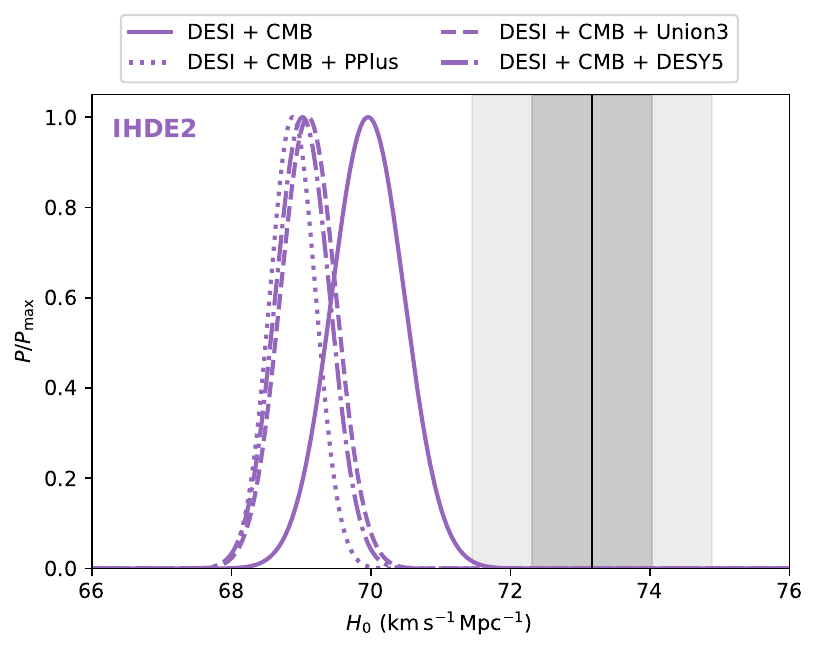}

    \caption{ One-dimensional marginalized posterior distributions for $H_0$ in the OHDE, BHDE, and IHDE2 models. Solid curves denote the $\textbf{DESI+CMB}$ dataset, while the dotted, dashed and dahs-dotted curves correspond to three different data combinations \textemdash{} $\textbf{DESI+CMB+PPlus}$, $\textbf{DESI+CMB+Union3}$, $\textbf{DESI+CMB+DESY5}$, respectively. }
    \label{fig.SN_comparison}
\end{figure}

As shown in the previous subsection, only models that adopt the future event horizon as the IR cutoff (OHDE, BHDE, IHDE2) can alleviate the Hubble tension problem. 
In order to test the robustness of this conclusion, in this subsection, we now examine whether it holds true for the cases of using alternative BAO and SN data.
Here we consider two sets of comparisons: (1) replacing DESI BAO data with the non-DESI compilation, combined with CMB data, and (2) replacing DESY5 SN data with PantheonPlus or Union3 compilations, combined with DESI+CMB.
Results for the alternative BAO comparison are presented in Fig. \ref{fig.BAO_comparison} and Table \ref{tab:BAO_comparison}. 
Results for the alternative SN comparison are presented in Fig. \ref{fig.SN_comparison} and Table \ref{tab:SN_comparison}.

Figure \ref{fig.BAO_comparison} compares the marginalized posterior distributions of $H_0$ for the OHDE, BHDE, and IHDE2 models using DESI+CMB (solid curves) and non-DESI+CMB (dashed curves) datasets.
Results for the OHDE model are shown in the top panel (orange), for the BHDE model in the bottom-left (pink), and for the IHDE2 model in the bottom-right (purple).
When DESI data are replaced with non-DESI data, the mean values of $H_0$ for the OHDE and BHDE models remain close to the SH0ES measurement, giving $H_0 = 71.39 \pm 1.65$ km s$^{-1}$ Mpc$^{-1}$ and $H_0 = 70.50 \pm 1.39$ km s$^{-1}$ Mpc$^{-1}$, respectively.
However, the mean value for the IHDE2 model decreases slightly, giving $H_0 = 68.98 \pm 1.67$ km s$^{-1}$ Mpc$^{-1}$.
More importantly, all three models exhibit noticeably broader posterior distributions, reflecting larger parameter uncertainties in $H_0$. 
As shown in Table \ref{tab:BAO_comparison}, the Hubble tension for the non-DESI+CMB dataset is reduced to $0.95\sigma$, $1.63\sigma$, and $2.23\sigma$ for the OHDE, BHDE, and IHDE2 models, respectively\footnote{\cite{Colgain:2021beg} argue that the ability of OHDE to resolve H0 tension can be traced to the presence of a turning point in $H(z)$, a sign of a violation of the null energy condition in the full model.}. 
Therefore, on the one hand, for HDE models that adopt the future event horizon as the IR cutoff, replacing DESI with the non-DESI data can also alleviate the Hubble tension;
on the other hand, since the DESI data reduce the error bar on $H_0$, it reveals a more pronounced Hubble tension problem.

\begin{table}
    \centering
    \begin{tabular}{lc|c}
    \hline \hline
      & \multicolumn{1}{c}{non-DESI+CMB}   & \multicolumn{1}{|c}{DESI+CMB}   \\
    \cline{2-3}
      \textbf{Model}    & $H_0$ tension  & $H_0$ tension \\
     \hline 
      OHDE          & 0.95$\sigma$   & 1.74$\sigma$ \\
      BHDE          & 1.63$\sigma$   & 2.65$\sigma$  \\
      IHDE2         & 2.23$\sigma$   & 3.19$\sigma$  \\
     \hline \hline \\[-1mm]
    \end{tabular}
    \\
    \begin{flushleft}
    \caption{Statistical significance of the Hubble tension derived from non-DESI+CMB and DESI+CMB datasets for the OHDE, BHD, and IHDE2 models.}
    \label{tab:BAO_comparison}
    \end{flushleft}
\end{table}

In Fig. \ref{fig.SN_comparison}, we explore the effects of incorporating different SN compilations into the DESI+CMB dataset, where results are presented in dotted, dashed, and dash-dotted lines for PantheonPlus, Union3, and DESY5, respectively.
Across all three models, adding SN data generally shifts the mean $H_0$ toward lower values compared to DESI+CMB alone. 
The inclusion of PantheonPlus data results in the lowest mean values: $H_0 = 69.72 \pm 0.32$ km s$^{-1}$ Mpc$^{-1}$ (the OHDE model), $H_0 = 69.51 \pm 0.30$ km s$^{-1}$ Mpc$^{-1}$ (the BHDE model), and $H_0 = 68.88 \pm 0.33$ km s$^{-1}$ Mpc$^{-1}$ (the IHDE2 model).
The inclusion of Union3 data gives slightly higher values: $H_0 = 70.27 \pm 0.37$ km s$^{-1}$ Mpc$^{-1}$ (the OHDE model), $H_0 = 69.92 \pm 0.34$ km s$^{-1}$ Mpc$^{-1}$ (the BHDE model), and $H_0 = 69.10 \pm 0.40$ km s$^{-1}$ Mpc$^{-1}$ (the IHDE2 model).
As shown in Table \ref{tab:SN_comparison}, compared with the DESI+CMB, incorporating PantheonPlus data increases the Hubble tension to $3.76\sigma$, 4.01$\sigma$, and 4.65$\sigma$ for the OHDE, BHDE, and IHDE2 models, respectively. 
Incorporating Union3 data gives tensions of $3.09\sigma$, 3.51$\sigma$, and 4.29$\sigma$ for the OHDE, BHDE, and IHDE2 models, respectively. 
Therefore, incorporating SN data into the DESI+CMB combination intensifies the Hubble tension problem.
In particular, the combination of DESI+CMB+PPlus gives the strongest tension, indicating that the Hubble tension problem is sensitive to the choice of SN compilation.
Overall, although incorporating SN data into the DESI+CMB increases the tensions, all three HDE models exhibit lower tensions than the $\Lambda$CDM, confirming that our conclusion remains robust across different SN datasets.

\begin{table*}
    \centering
    \begin{tabular}{lccc|c}
    \hline \hline
     & \multicolumn{1}{c}{DESI+CMB+PPlus}   & \multicolumn{1}{c}{DESI+CMB+Union3}   & \multicolumn{1}{c}{DESI+CMB+DESY5} & \multicolumn{1}{|c}{DESI+CMB}  \\
    \cline{2-5}
      \textbf{Model}   & $H_0$ tension  & $H_0$ tension  & $H_0$ tension  & $H_0$ tension \\
     \hline
      OHDE          & 3.76$\sigma$  &  $3.09\sigma$ & 2.49$\sigma$  & 1.74$\sigma$  \\
      BHDE          & 4.01$\sigma$  &  $3.51\sigma$ & 3.12$\sigma$  & 2.65$\sigma$  \\
      IHDE2         & 4.65$\sigma$  &  $4.29\sigma$ & 4.49$\sigma$  & 3.19$\sigma$  \\
     \hline \hline
    \end{tabular}
    \begin{flushleft}
    \caption{Statistical significance of the Hubble tension derived from the DESI+CMB+PPlus, DESI+CMB+Union3, DESI+CMB+DESY5, and DESI+CMB datasets for the OHDE, BHD, and IHDE2 models.}
    \label{tab:SN_comparison}
    \end{flushleft}
\end{table*}

One of the main conclusions of this paper is that HDE models using the future event horizon as the IR cutoff can partially mitigate the Hubble tension, whereas those based on the Hubble scale (or its extensions) cannot.
From a physical perspective, the choice of IR cutoff can significantly affect the late-time evolution of the EoS parameter $w(z)$ of HDE.
For HDE models employing the Hubble scale or its extensions, $w(z)$ evolves from $w>-1$ at early times to $w \approx -1$ at late times, so the model behavior is similar to $\Lambda$CDM at late times and gives a low $H_0$.
In contrast, for HDE models employing the future event horizon, $w(z)$ evolves from $w>-1$ at early times to $w<-1$ at late times.
Therefore, the choice of the future event horizon leads to a more dynamical evolution of $w(z)$ at late times, which enlarges the parameter space and introduces the parameter degeneracy.
Consequently, the degeneracy between $H_0$ and $w(z)$ allows for a higher mean value of $H_0$ and a larger error bar, thereby reducing the Hubble tension.
However, the DESI analysis \citep{DESI:2025fii} suggests that $w(z)$ should evolve from $w<-1$ to $w>-1$.
Therefore, the choice of future event horizon in HDE can raise the value of $H_0$, but at the cost of a poor statistical preference.

\section{Summary}
\label{sec:conclusion}
In this paper, we systematically investigate the Hubble tension problem in the framework of HDE. 
We select six representative theoretical models from all four categories of HDE, including OHDE, GRDE, IHDE1, IHDE2, THDE, and BHDE. To obtain cosmological constraints, we adopt the latest BAO data from DESI DR2, CMB distance priors from $Planck$ 2018, and the DESY5 compilation of SN data. In addition, we also consider alternative datasets including the PantheonPlus and Union3 SN compilations, as well as a collection of alternative BAO data, including 6dFGS+SDSS MGS, BOSS Galaxy, eBOSS LRG, DES Y3, eBOSS Quasar, and eBOSS Ly$\alpha$-forest surveys.

By analyzing results based on the DESI+CMB and DESI+CMB+DESY5 datasets, we obtain two key conclusions. 
First, HDE models that employ the Hubble scale or its combinations as the IR cutoff cannot alleviate the Hubble tension problem, maintaining tensions of 5$\sigma$-6$\sigma$ that are comparable to the $\Lambda$CDM.
Second, HDE models that employ the future event horizon as the IR cutoff can partially mitigate the Hubble tension problem, reducing the tensions to below 3.19$\sigma$ for DESI+CMB and below 4.49$\sigma$ for DESI+CMB+DESY5 (compared to over 5.4$\sigma$ for the $\Lambda$CDM model).

To explore whether our key conclusions hold true for other BAO data and SN data, we also take into account alternative data combinations, including non-DESI+CMB, DESI+CMB+PPlus, and DESI+CMB+Union3. 
When replacing DESI with non-DESI data, HDE models that employ the future event horizon as the IR cutoff still have the ability to alleviate the Hubble tension problem, with tension reduced to below 2.23$\sigma$.
Although incorporating SN data into the DESI+CMB combination increases the tension compared to DESI+CMB alone, HDE models that employ the future event horizon still exhibit lower tensions (below 4.65$\sigma$) than the $\Lambda$CDM model.
Therefore, our key conclusions are insensitive to the cases of using alternative BAO data and SN data.

Current cosmological measurements indicate a Hubble tension that is larger than 5$\sigma$ for the $\Lambda$CDM model.
As shown in this paper, some HDE models that based on the future event horizon can partially mitigate the Hubble tension problem. 
However, these HDE models perform poorer than the $\Lambda$CDM model in cosmology-fits \citep{Li:2024bwr}.
Therefore, it is interesting to search for some other dynamical DE models that can genuinely resolve the Hubble tension problem and give a good cosmological fitting result that is comparable to the $\Lambda$CDM model.
In addition, since the Hubble tension problem primarily arises from a discrepancy between high-redshift probes (CMB) and low-redshift probes (Cepheids and SNe), incorporating some other cosmological probes such as standard sirens (gravitational waves) may have the potential of alleviating the Hubble tension problem.
These open questions merit further investigation.

\section*{Acknowledgements}
The authors thank the referee for many important comments that improved the manuscript. This work was supported by a grant of Guangdong Provincial Department of Science and Technology under No. 2020A1414040009.


\section*{Data Availability}



All the data used are explained in the text and are publicly available.



\bibliographystyle{mnras}
\bibliography{example} 







\bsp	
\label{lastpage}
\end{document}